\documentclass[letterpaper,twocolumn,10pt]{article}
\usepackage{zhanggroup}

\usepackage{hyperref}
\usepackage{url}
\usepackage{graphicx}
\usepackage{xspace}
\usepackage{booktabs}
\usepackage{makecell}
\usepackage{multirow}
\usepackage[font=footnotesize]{subcaption}
\usepackage{wrapfig}
\usepackage{xcolor}
\usepackage{mdframed}
\usepackage{tikz}
\usetikzlibrary{positioning}
\usetikzlibrary{shapes.geometric}
\usetikzlibrary{arrows.meta}
\usepackage{amsmath} 

\newcommand{\customTableFont}{\fontsize{8pt}{8pt}\selectfont}
\newcommand{\hytt}[1]{\texttt{\hyphenchar\font=\defaulthyphenchar #1}}
\newcommand{\hyit}[1]{\textit{\hyphenchar\font=\defaulthyphenchar #1}}
\newcommand{\hybf}[1]{\textbf{\hyphenchar\font=\defaulthyphenchar #1}}
\newcommand{\mypara}[1]{\noindent\textbf{{{#1.\xspace}}}}
\newcommand{\refappendix}[1]{\hyperref[#1]{Appendix~\ref*{#1}}}

\begin{document}

\date{}

\title{\bf Benchmark of Benchmarks: Unpacking Influence and Code Repository Quality in LLM Safety Benchmarks}

\author{
Junjie Chu\textsuperscript{1}\ \ \
Xinyue Shen\textsuperscript{1,2}\ \ \
Ye Leng\textsuperscript{1}\ \ \
Michael Backes\textsuperscript{1}\ \ \
Yun Shen\textsuperscript{3}\ \ \
Yang Zhang\textsuperscript{1}\textsuperscript{$\clubsuit$}\ \ \
\\
\\
\textsuperscript{1}\textit{CISPA Helmholtz Center for Information Security} \ \ \ 
\textsuperscript{2}\textit{University of Waterloo} \ \ \
\textsuperscript{3}\textit{Flexera} \ \ \
}

\maketitle
\def\thefootnote{$\clubsuit$}\footnotetext{Corresponding author.}\def\thefootnote{\arabic{footnote}}

\begin{abstract}
The rapid expansion of research in LLM safety presents challenges in tracking advancements, making benchmarks important evaluation infrastructures for identifying key trends and facilitating systematic comparisons.
Yet no systematic assessment exists of their code quality and runnability, nor of what factors are associated with the community's adoption of certain benchmarks over others.
To address this gap, we conduct a systematic measurement study of 31 LLM safety benchmarks (covering prompt injection, jailbreak, and hallucination) with 382 non-benchmark papers as a control group, combining automated static analysis, human runnability testing (220+ person-hours), and bibliometric analysis.
We find that only 39\% of benchmark repositories can run without modification, only 16\% provide flawless installation guides, and a mere 6\% include ethical considerations despite containing potentially harmful content.
These deficiencies persist across the study period with no significant improvement.
Analyzing adoption factors, we find that benchmark adoption correlates with author prominence and code runnability, but not with code quality standards such as Pylint score and maintainability, suggesting that the community's benchmark selection does not reward higher coding standards.
Based on these results, we identify potential safety and reliability concerns.
Some safety benchmark repositories openly expose harmful content, such as successful jailbreak responses, without any ethical warning or access control, effectively serving as unguarded attack resources.
Furthermore, when benchmarks require ad-hoc modifications to run, downstream safety evaluations across different papers may not be comparable.
We present case studies illustrating these concrete consequences and propose a targeted checklist to help benchmark contributors improve code quality, documentation, and ethical practices.
We hope our work contributes to improving the benchmark code quality and mitigating the related security and reliability concerns.
\end{abstract}

\section{Introduction}
\label{section:introduction}

\begin{figure}[!t]
\centering
\includegraphics[width=0.618\columnwidth]{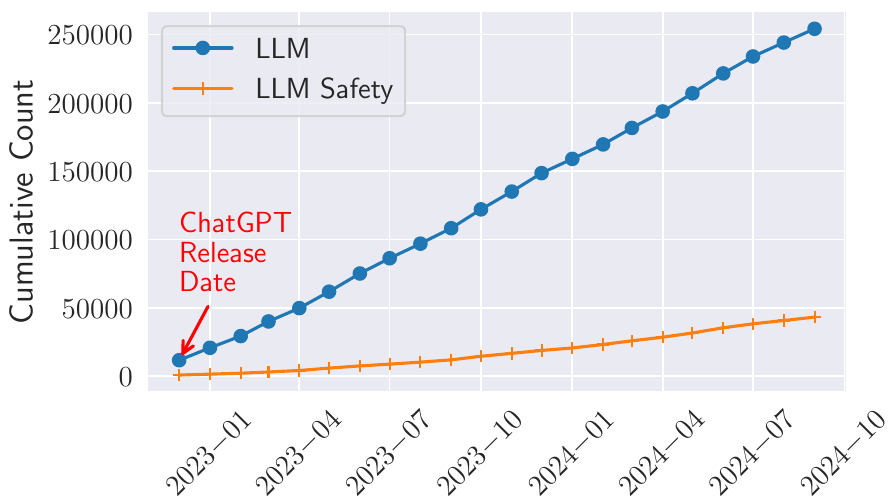}
\caption{
Cumulative count of LLM and LLM safety papers since ChatGPT was released (data from Semantic Scholar).
}
\label{figure:cumulative_papers_count}
\end{figure}

LLM safety is experiencing an arms race, with new attack techniques and defensive measures~\cite{RSD23, JLFYSXIBMF23, LCZNW23, CLYSBZ25, PR22, GAMEHF23,TMSAABBBBBBBCCCEFFFFGGGHHHIKKKKKKLLLLLMMMMMNPRRSSSSSTTTWKXYZZFKNRSES23, OWJAWMZASRSHKMSAWCLL22,CLYLLSBSZ25} being proposed constantly. 
Within two years of ChatGPT's launch~\cite{chatgpt_launch}, this dynamic has spurred the publication of over 250,000 new papers about LLMs, with nearly 50,000 on safety (see~\autoref{figure:cumulative_papers_count}).

The rapid growth of research in this area brings challenges in tracking the latest advancements.
In this context, benchmark projects serve as an increasingly important evaluation infrastructure by comparing research findings and providing valuable insights.
As such, the code quality of these benchmarks matters: if they suffer from poor code quality, the safety evaluations built on top of them may be compromised.
This raises a natural follow-up question: Does the community tend to adopt benchmarks with higher code quality, or are other factors at play?
If widely adopted benchmarks suffer from code deficiencies, researchers across the community would inevitably spend additional time debugging and modifying the code before they can conduct their evaluations, and the resulting evaluation outcomes may vary depending on how each researcher resolves the issues.
Yet, no systematic assessment has been conducted on the code quality of LLM safety benchmarks, nor on what factors are associated with their adoption.

\mypara{Research Questions}
To address these gaps, we formulate two research questions (RQs):
\begin{itemize}
    \item \textbf{RQ1:} What is the current state of code quality and runnability in LLM safety benchmark repositories?
    \item \textbf{RQ2:} Which factors are associated with the community’s adoption of benchmarks? In particular, is adoption related to code quality or to other external factors?
\end{itemize}

\mypara{Approach and Findings}
To answer the RQs, we first build a dataset covering the period from November 30, 2022 (ChatGPT's launch) to November 1, 2024.
The dataset comprises 31 LLM safety benchmark papers (with 27 public repositories) and 382 non-benchmark papers (the control group, with 168 public repositories) across three key newly emerging topics: prompt injection, jailbreak, and hallucination.
Notably, the 31 benchmarks represent a census of all qualifying benchmarks within the study period, not a sample drawn from a larger population.

We then conduct a systematic measurement study and statistical analysis based on the collected data.
Specifically, we first assess the code quality and runnability of benchmark repositories using both tool-based methods (Pylint and Radon) and human-based runnability testing on a dedicated server (Ubuntu 20.04, 4$\times$A100 GPUs), requiring over 220 person-hours.
Then, we investigate what factors correlate with benchmark adoption, using citation density as a proxy for adoption and examining both internal code properties (runnability, static quality metrics, maintenance frequency) and external factors (author prominence, institutional affiliation, geolocation, publication status).

For \textbf{RQ1}, we find that only 39\% of benchmark repositories can run without modification, only 16\% provide flawless installation guides, and only 6\% include ethical considerations.
These deficiencies persist across the study period with no significant temporal improvement.
Overall, \emph{the code quality and supplementary material quality of LLM safety benchmarks have substantial room for improvement}.

For \textbf{RQ2}, we find that benchmark adoption is associated with author prominence and code runnability, but not with code quality standards.
Author prominence correlates with citation density but shows no significant correlation with code quality.
Ready-to-use code significantly relates to higher citation density ($p=0.004$), while code requiring modifications does not.\footnote{The threshold we use for related adjusted p-values ($p$) is 0.05~\cite{F70}.}
In other words, \emph{the community's benchmark selection does not reward higher coding standards}.

\mypara{How the LLM Safety Ecosystem Is Affected}
Based on these findings, we identify potential \textbf{safety and reliability concerns} in the current LLM safety ecosystem.
On the safety side, some benchmark repositories openly expose harmful content (e.g., successful jailbreak responses) without any ethical warning or access control, effectively serving as unguarded attack resources.
On the reliability side, when benchmarks lack proper dependency management (e.g., relying on the OpenAI Python library without specifying a version), breaking changes such as the \texttt{openai} v1.0 migration can cause execution failures, forcing different users to apply different ad-hoc fixes to make the code runnable, which may lead to divergent evaluation results across papers.
We further present case studies to substantiate the existence and scope of both concerns in current benchmark repositories.
In addition, we provide a checklist covering code quality, dependency management, documentation, and ethical practices, which we hope will help benchmark contributors improve the quality and usability of their repositories.

\mypara{Main Contributions}
We summarize our main contributions as follows:
\begin{itemize}
\item
We conduct the first systematic measurement study and statistical analysis of LLM safety benchmark repositories, combining automated static analysis, human runnability testing and bibliometric analysis.
\item
We reveal significant code quality and documentation deficiencies in existing benchmarks and show that these deficiencies persist over time. 
\item
We analyze the factors associated with benchmark adoption and uncover a potential misalignment: the community's adoption correlates with author prominence and code runnability, but not with code quality standards.
\item
We identify safety and reliability concerns arising from the current benchmark ecosystem, including the unguarded exposure of harmful content and the potential for divergent evaluations due to dependency mismanagement.
\end{itemize}

\section{Data Under Study}
\label{section:data}

\subsection{Scope of This Study}

In this paper, we focus on three novel safety risks specific to LLMs, including \hybf{prompt injection}~\cite{PR22, GAMEHF23}, \hybf{jailbreak}~\cite{ZWKF23, SCBZ23, CLYSBZ25}, and \hybf{hallucination}~\cite{RSD23, JLFYSXIBMF23, LCZNW23}.
These newly emerging safety topics in the LLM era pose common and serious real-world risks, offering us unique research opportunities.
More details about why these topics are chosen are in~\refappendix{section:intro_topic}.

\subsection{Data Collection}
\label{section:paper_data_collection}

\begin{figure}[!t]
    \centering
    \includegraphics[width=0.45\textwidth]{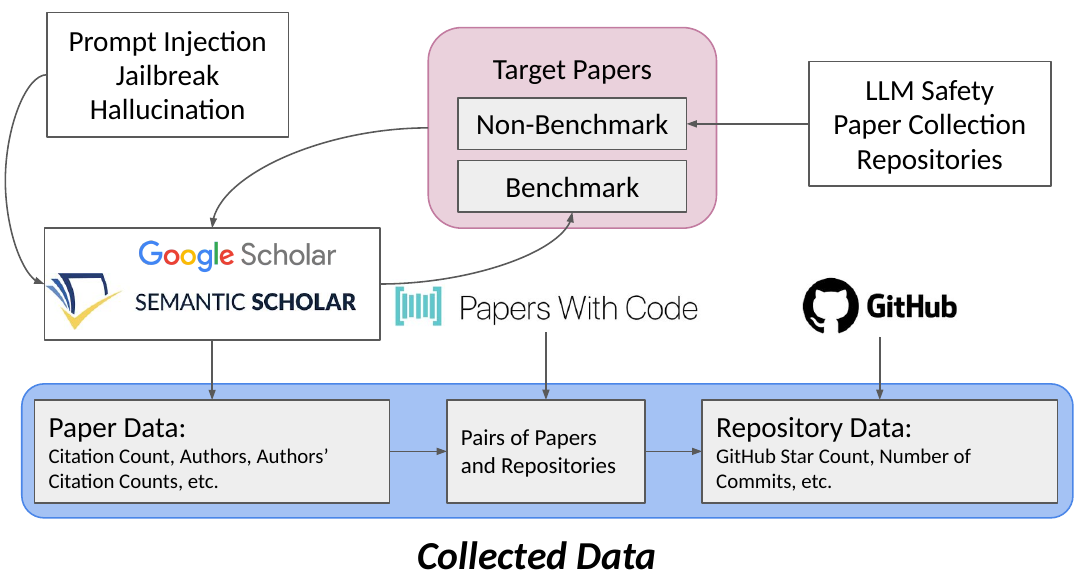}
    \caption{Data collection pipeline.}
    \label{figure:data_workflow}
\end{figure}

\mypara{Data Sources}
The metadata of papers is sourced from Semantic Scholar~\cite{semantic_scholar_api} and Google Scholar~\cite{google_scholar}.
We retrieve the paper-repository pairs from Papers with Code~\cite{paper_with_code_api}.
The metadata related to repositories, such as the GitHub star count, is collected from GitHub API~\cite{github_api}.
The overall data collection workflow is illustrated in~\autoref{figure:data_workflow}.
Details of these data sources are introduced in~\refappendix{section:data_source_details} and the details of metadata are available in~\refappendix{section:metadata_detail}.
The time period we study is from November 30, 2022 (ChatGPT's launch) to November 1, 2024.

\mypara{Benchmark Papers}
In order to conduct a comprehensive compilation of benchmark papers across selected safety topics, we implement a systematic collection methodology utilizing three distinct keyword sets (detailed in~\refappendix{section:keywords}). 
Our data collection process leverages both the Semantic Scholar API and the Google Scholar platform to attempt comprehensive coverage of the literature on LLM safety.
For each safety topic, we first query the Semantic Scholar API with the corresponding keyword set to search for related papers, targeting both titles and abstracts.
The results are then complemented by direct searches using the same keyword sets on the Google Scholar website.
Notably, our findings reveal a substantial overlap between results from both platforms (only two papers from each source that do not overlap), aligning with those from previous studies~\cite{H21}.
We combine all overlapping and non-overlapping papers (in total 39 papers) for further inspection.

Subsequently, we conduct a rigorous manual inspection process to check if the collected benchmark papers meet both relevancy and quality criteria. 
Specifically, we filter papers according to the following three criteria:
\hybf{[1]} The paper must be relevant to LLM safety and should focus on the selected safety topic, rather than matching keywords in unrelated fields (e.g., ``iOS jailbreak'').
\hybf{[2]} The paper should not be a SoK or survey paper; we retain only primary research papers rather than higher-level overview or position works.
\hybf{[3]} The paper must be a genuine benchmark study, as evidenced by its research content, rather than merely referencing benchmark datasets in its methodology (e.g., ``We evaluate our method on benchmark datasets.'').
The PRISMA diagram is shown in~\autoref{figure:prisma} of~\refappendix{section:prisma}.
\hyit{After the inspection, we keep 31 inspected benchmark papers.}
As an additional completeness check, we performed supplementary searches using alternative terms (``testbed'' and ``challenge set''); no additional LLM safety benchmarks meeting our inclusion criteria were found.

\mypara{Non-Benchmark Papers}
To facilitate comparative analysis and assess the adoption of benchmark papers, we additionally collect non-benchmark papers within the same safety topics to serve as control groups.
We initially experimented with the Semantic Scholar API and the Google Scholar website to collect papers that are not benchmarks. 
However, such an attempt yielded an excessive volume of results, with most having a low relevance.
For example, when trying to retrieve non-benchmark papers about hallucination, we obtain about 37,100 papers from Semantic Scholar, while most of them only mention ``hallucination'' in the abstract as background knowledge.
Consequently, we refine our method and switch to examining four actively maintained GitHub repositories specializing in LLM safety literature~\cite{llm_ssp_1,llm_ssp_2,llm_ssp_3,llm_ssp_4}.\footnote{We manually verified that all collected benchmark papers were also present in the four repositories, with no additional benchmark papers identified. Thus, benchmark and non-benchmark papers' data sources can be considered similar.}
Specifically, we manually identify LLM safety-related papers from the above repositories, adhering to two main criteria: 
\hybf{[1]} Papers must fall within the three selected topics. 
\hybf{[2]} Papers are not benchmark, SoK, or survey papers.
\hyit{Finally, we keep 382 screened non-benchmark papers.}
Importantly, since all 31 benchmark papers also appear in the same four curated repositories (as verified in the footnote above), any curation bias affects both groups symmetrically.
Our findings therefore characterize differences \emph{within} high-visibility LLM safety research.
Moreover, if the control group were replaced with average-quality papers (with lower citations), benchmarks would appear relatively more influential, making our comparison conservative.

\mypara{Associated GitHub Repositories}
We employ the Papers with Code API to search for the official code repositories corresponding to each paper.
To the best of our knowledge, Papers with Code is the only systematic tool for connecting papers with code repositories.
We match only repositories marked as ``official,'' which means the repositories were provided by the authors of the corresponding papers.
We then conduct manual verification to exclude any potential erroneous pairs or empty repositories.
We find that the results from Papers with Code are highly accurate, with only three errors detected in 198 pairs. 
This may be because Papers with Code has already performed manual verification.

\mypara{Summary of the Collected Data}
We summarize the collected data in~\autoref{table:summary_datasets}.

\begin{table}[!t]
    \centering
    \caption{Summary of the Data Under Study}
    \label{table:summary_datasets}
    \scalebox{0.77}{
    \setlength{\tabcolsep}{3pt}
    \customTableFont 
    \begin{tabular}{lcc}
        \toprule
        \textbf{} & \textbf{Benchmark} & \textbf{Non-Benchmark} \\
        \midrule
        \textbf{Collection Period} & \multicolumn{2}{c}{[November 30, 2022 -- November 1, 2024]} \\
        \textbf{\# Papers} & 31   & 382  \\
        \textbf{\# Repositories} & 27 & 168  \\
        \textbf{\# Jailbreak} & 12 & 246  \\
        \textbf{\# Hallucination} & 13 & 103  \\
        \textbf{\# Prompt Injection} & 6 & 33   \\
        \bottomrule
    \end{tabular}
    }
\end{table}

\section{Code Quality and Runnability of LLM Safety Benchmarks (RQ1)}
\label{section:quality_repository}

We assess the code quality and runnability of benchmark repositories using automated tools, with non-benchmark papers serving as a control group for comparison.
We then complement the tool-based evaluation with human-based evaluation, which involves manually running the code and assessing the quality of supplementary materials, such as installation guides, data guides, and ethical considerations.

\subsection{Methodology of Tool-Based Evaluation}
\label{section:tool_based}

Python is the dominant language for LLM-related libraries, and all collected repositories are primarily built on Python.
Thus, we use two popular static analysis tools, Pylint~\cite{pylint} and Radon~\cite{radon}, to assess code quality.
We compute four metrics: \hyit{Pylint Score}, \hyit{Cyclomatic Complexity}, \hyit{Maintainability Index}, and \hyit{Number of Static Errors}.
Also, we evaluate repository maintenance using four GitHub API-based metrics: \hyit{Reply Time}, \hyit{Last Commit Time}, \hyit{Number of Commits}, and \hyit{Commit Frequency}.
These eight metrics collectively provide a view of both code quality and maintenance status (definitions in~\refappendix{section:quality_metric_details}).
We adopt non-benchmark papers in the same safety topics as the control group for comparison.

\subsection{Methodology of Human-Based Evaluation}
\label{section:human_based}

Tool-based evaluation has limitations, such as missing dynamic behavior detection and potential false negatives.
To address this, we complement it with human-based evaluation.

For available repositories, we clone them and execute all code on our server (Ubuntu 20.04, four A100 GPUs).
Following the README instructions, we run example scripts under recommended settings, requesting necessary access when required.
If the code remains unrunnable after four hours of setup and debugging, we mark it as \hybf{not runnable}.
We log any additional modifications (e.g., bug fixes) beyond those in the guides.
If scripts run successfully, we record execution time.
Given LLMs' high computational demands and the unpredictability of external LLM APIs, we set a four-hour time limit, exceeding that in previous studies~\cite{CP16,OLSWKPUBT23}.
We also manually evaluate repositories' supplementary materials, as they significantly impact usability.
First, we assess the quality of install guides, which directly influence users' success and efficiency in running the code.
Next, we examine the presence of data guides, which help users organize their data and execute the code effectively.
Finally, we check for ethical considerations, ensuring repositories provide guidance for responsible usage, especially in LLM safety contexts where they may contain or generate harmful content.
For each available code repository, we run the experimental artifact twice.
Two doctoral researchers with varying levels of experience in LLMs (one junior and one senior) have conducted the two attempts independently.
Our manual evaluation requires over 220 person-hours.
We accept the most positive results for the repositories as the previous work~\cite{OLSWKPUBT23}.
The measured metrics are summarized in~\autoref{table:human_metrics}.

\begin{table}[!ht]
\centering
\caption{An outline of the metrics measured during the human-based evaluation.}
\label{table:human_metrics}
\scalebox{0.77}{
    \setlength{\tabcolsep}{3pt}
    \customTableFont
    \begin{tabular}{m{0.2\columnwidth}|m{0.6\columnwidth}}
    \toprule
    \makecell{\textbf{Aspect}} & \makecell{\textbf{Description}} \\
    \midrule
    \makecell{\multirow{3}[3]{*}[-0ex]{\makecell{Code\\Quality}}} & Is the necessary dataset available in the repository? \\
          & Are the provided example scripts runnable? \\
          & Are there any extra modifications required (such as fixing bugs) besides those mentioned in the guides when trying to run the example scripts? \\
    \midrule
    \makecell{\multirow{3}[3]{*}[1ex]{\makecell{Supplementary\\Material Quality}}} & Does the repository provide good install guides? \\
          & Does the repository provide useful data guides? \\
          & Does the repository contain essential ethical considerations? \\
    \bottomrule
    \end{tabular}
    }
\end{table}

\mypara{Inter-Rater Reliability}
We report the inter-rater agreement for all five human evaluation dimensions in~\autoref{table:inter_rater}.
Three dimensions (Ethical Consideration, Data Available, and Data Guidance) achieve perfect agreement ($\kappa=1.0$).
For Runnable Demo, Cohen's $\kappa=0.704$ (substantial) and Gwet's AC1$=0.885$ (almost perfect) confirm high agreement.
For Install Guide, Cohen's $\kappa=0.467$ is deflated by the prevalence paradox~\cite{G08}: 22 repositories were rated ``Yes'' by both evaluators, leaving only 2 borderline cases.
Gwet's AC1, which is robust to prevalence imbalance, yields 0.902 (almost perfect).

\begin{table}[!t]
    \centering
    \caption{Inter-rater agreement for human-based evaluation.}
    \label{table:inter_rater}
    \scalebox{0.77}{
    \setlength{\tabcolsep}{3pt}
    \customTableFont
    \begin{tabular}{cccc}
    \toprule
    \textbf{Dimension} & \textbf{Cohen's $\kappa$} & \textbf{Gwet's AC1} & \textbf{PABAK} \\
    \midrule
    Runnable Demo & 0.704 & 0.885 & 0.833 \\
    Install Guide & 0.467 & 0.902 & 0.833 \\
    Ethical Consideration & 1.000 & 1.000 & 1.000 \\
    Data Available & 1.000 & 1.000 & 1.000 \\
    Data Guidance & 1.000 & 1.000 & 1.000 \\
    \bottomrule
    \end{tabular}
    }
\end{table}

\subsection{Tool-Based Results of Benchmark Papers}
\label{section:tool_results}

\mypara{Code Repository Availability}
Our measurement results (~\refappendix{section:supplementary_figures}) show that \hybf{benchmark papers better adhere to open science policies}.
Specifically, across all topics, 87\% of benchmark papers provide accessible repositories, compared to only 44\% of non-benchmark papers.
Examining each topic individually reveals a similar trend: for prompt injection, jailbreak, and hallucination, the proportion of benchmark papers offering available repositories reaches 100\%, 75\%, and 92\%, respectively, significantly higher than those of corresponding non-benchmark papers.

\begin{table}[!t]
    \centering
    \caption{
        Code quality M-W U test results with 95\% bootstrap CIs.
        Positive $\delta$: \hyit{non-benchmark} dominates; negative: \hyit{benchmark} dominates.
    }
    \label{table:mwu_cliff_code}
    \scalebox{0.72}{
    \setlength{\tabcolsep}{3pt}
    \customTableFont
    \begin{tabular}{ccccc}
    \toprule
    \textbf{Metric} & \textbf{$p$-value} & \textbf{Cliff's $\delta$} & \textbf{95\% CI} & \textbf{Effect Size} \\
    \midrule
    Pylint Score & 0.031 & $-$0.276 & [$-$0.525, $-$0.014] & Small \\
    Cyclomatic Complexity & 0.649 & $-$0.055 & [$-$0.296, 0.195] & Negligible \\
    Maintainability Index & 0.244 & $-$0.140 & [$-$0.388, 0.113] & Negligible \\
    Number of Static Errors & 0.783 & $-$0.014 & [$-$0.132, 0.078] & Negligible \\
    Reply Time (Hours) & 0.044 & $-$0.239 & [$-$0.426, $-$0.040] & Small \\
    Last Commit Time (Days) & 0.491 & 0.083 & [$-$0.138, 0.299] & Negligible \\
    Number of Commits & 0.001 & $-$0.389 & [$-$0.575, $-$0.193] & Medium \\
    Commit Frequency & 0.010 & $-$0.309 & [$-$0.516, $-$0.094] & Small \\
    \bottomrule
    \end{tabular}
    }
\end{table}

\mypara{Code Repository Quality}
We use non-parametric tests to statistically compare benchmark and non-benchmark papers across eight code quality metrics.\footnote{The descriptive statistic analysis of code repository quality (e.g., average values) is in~\refappendix{section:descriptive_code}.}
Following rules of thumb in Statistics~\cite{MV18,HT10,VM07}, we set a minimum sample size of 25 per group, requiring us to analyze all benchmark papers collectively (see~\autoref{table:summary_datasets}) rather than segmenting them by topic.
Notably, our 31 benchmark papers constitute a \emph{census} of all qualifying benchmarks within the study period, not a sample drawn from a larger population; this strengthens the representativeness of our dataset despite its modest size.
We apply the Mann-Whitney U (M-W U) test~\cite{MN102,N08} to assess statistical significance~\cite{H14,O63,WW17} and use Cliff's delta~\cite{HK04,MRL10,MY24} to measure practical significance (effect size)~\cite{penn_practical,P16,K96,PMO19}.\footnote{
Following Cliff's delta guidelines~\cite{HK04}, effect sizes are classified as: negligible (\(|\delta|\leq0.147\)), small (\(0.147<|\delta|\leq0.330\)), medium (\(0.330<|\delta|\leq0.474\)), and large (\(|\delta|>0.474\)).
}
Detailed results are in~\autoref{table:mwu_cliff_code}.
The tests show that benchmark papers dominate in Pylint score with a small effect size ($|\delta|=0.276$), suggesting better code quality.
Additionally, benchmark papers lead in three repository maintenance metrics (Reply Time, Number of Commits, and Commit Frequency), indicating that their authors are more active in maintaining and engaging with their repositories.
For other metrics, no significant distribution differences are observed.
In summary, we find that \textbf{benchmark papers exhibit higher code repository quality in terms of Pylint score and maintenance frequency}.
A leave-one-out sensitivity analysis shows that Number of Commits and Commit Frequency are fully stable (significant in all 31 iterations).
Pylint Score remains mostly stable (significant in 23/31 iterations, with $\delta$ consistently in [0.244, 0.327]).
However, Reply Time (original $p=0.044$) is sensitive to sample composition (significant in only 16/31 iterations), suggesting this borderline finding should be interpreted with caution.

\mypara{Statistical Robustness}
To assess whether the null results reflect genuine absence of effects rather than insufficient power, we conduct a sensitivity power analysis at a fixed target effect size ($|\delta|=0.33$, medium) following previous work~\cite{PGC18}, avoiding the known circularity of classical post-hoc power analysis~\cite{HH01}.
As shown in~\autoref{table:power_analysis_code}, for all four non-significant code quality metrics, power exceeds 0.82, above the conventional 0.80 threshold~\cite{C13}.
Combined with negligible observed effect sizes ($|\delta|=0.014$--$0.140$), these results confirm that the null findings represent genuinely small effects, not artifacts of insufficient sample size.
The bootstrap confidence intervals in~\autoref{table:mwu_cliff_code} corroborate this: for significant results, the CI excludes zero; for null results, the CI includes zero.

\begin{table}[!t]
    \centering
    \caption{
        Sensitivity power analysis for code quality metrics at fixed target $|\delta|=0.33$.
        Metrics marked with * are statistically significant ($p<0.05$).
    }
    \label{table:power_analysis_code}
    \scalebox{0.77}{
    \setlength{\tabcolsep}{3pt}
    \customTableFont
    \begin{tabular}{cccc}
    \toprule
    \textbf{Metric} & $|\delta|_{\text{obs}}$ & \textbf{$p$-value} & \textbf{Power} \\
    \midrule
    Pylint Score* & 0.276 & 0.031 & 0.774 \\
    Cyclomatic Complexity & 0.055 & 0.649 & 0.822 \\
    Maintainability Index & 0.140 & 0.244 & 0.822 \\
    Number of Static Errors & 0.014 & 0.783 & 0.822 \\
    Reply Time (Hours)* & 0.239 & 0.044 & 0.822 \\
    Last Commit Time (Days) & 0.083 & 0.491 & 0.822 \\
    Number of Commits* & 0.389 & 0.001 & 0.822 \\
    Commit Frequency* & 0.309 & 0.010 & 0.822 \\
    \bottomrule
    \end{tabular}
    }
\end{table}

\subsection{Human-Based Results of Benchmark Papers}
\label{section:human_evaluation_results}

\begin{figure*}[!t]
\centering
\begin{subfigure}{0.32\textwidth}
\centering
\includegraphics[width=\linewidth]{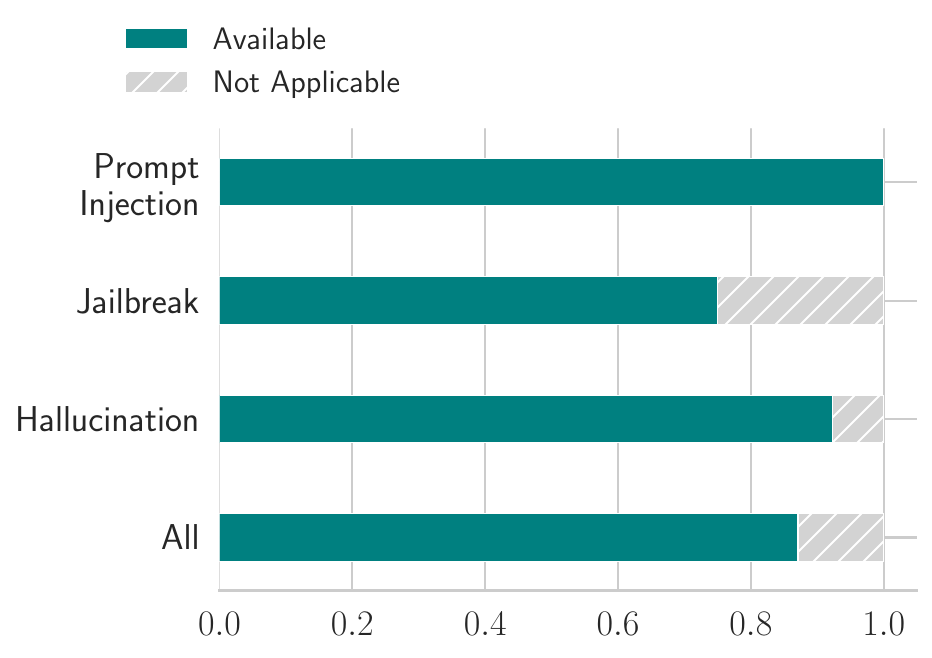}
\caption{
Data availability status.
}
\label{figure:rose_data}
\end{subfigure}
\begin{subfigure}{0.32\textwidth}
\centering
\includegraphics[width=\linewidth]{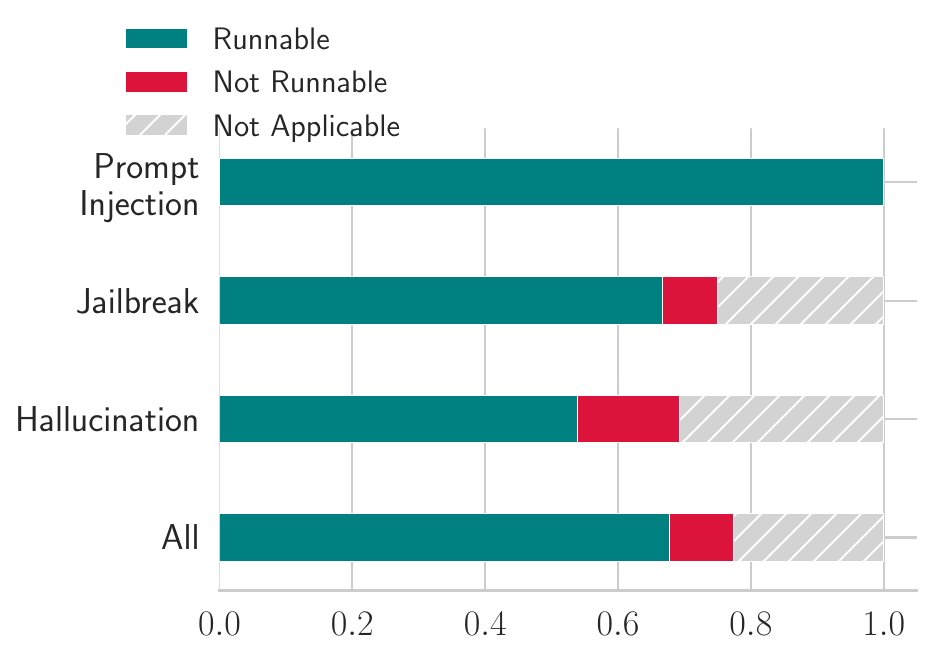}
\caption{
Runnable code status.
}
\label{figure:rose_runnable_demo}
\end{subfigure}
\begin{subfigure}{0.32\textwidth}
\centering
\includegraphics[width=\linewidth]{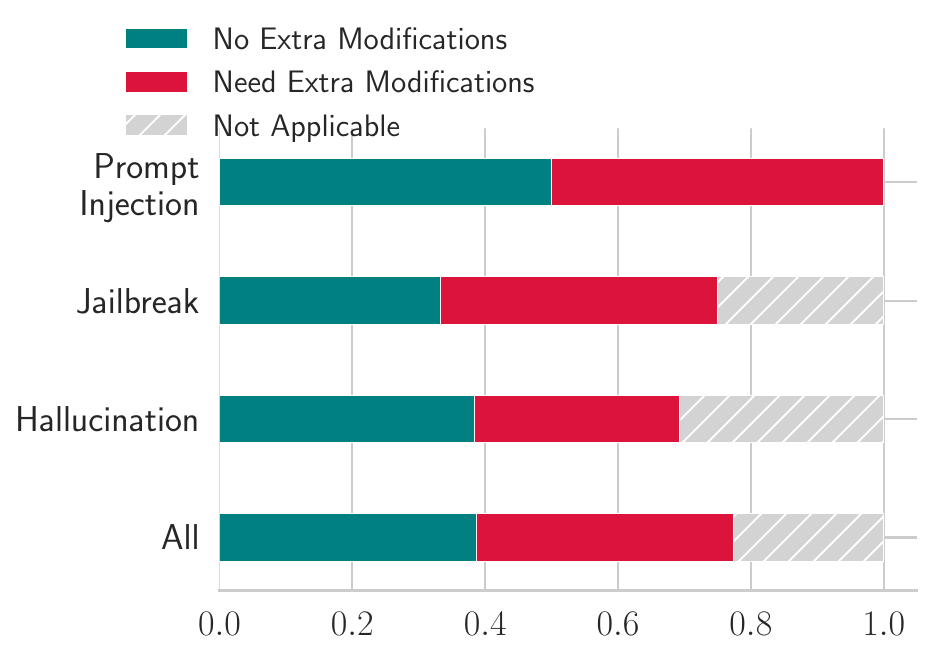}
\caption{
Extra modification status.
}
\label{figure:rose_bugs}
\end{subfigure}
\caption{
Human-based evaluation results of code quality.
}
\end{figure*}

\begin{figure*}[!t]
\centering
\begin{subfigure}{0.32\textwidth}
\centering
\includegraphics[width=\linewidth]{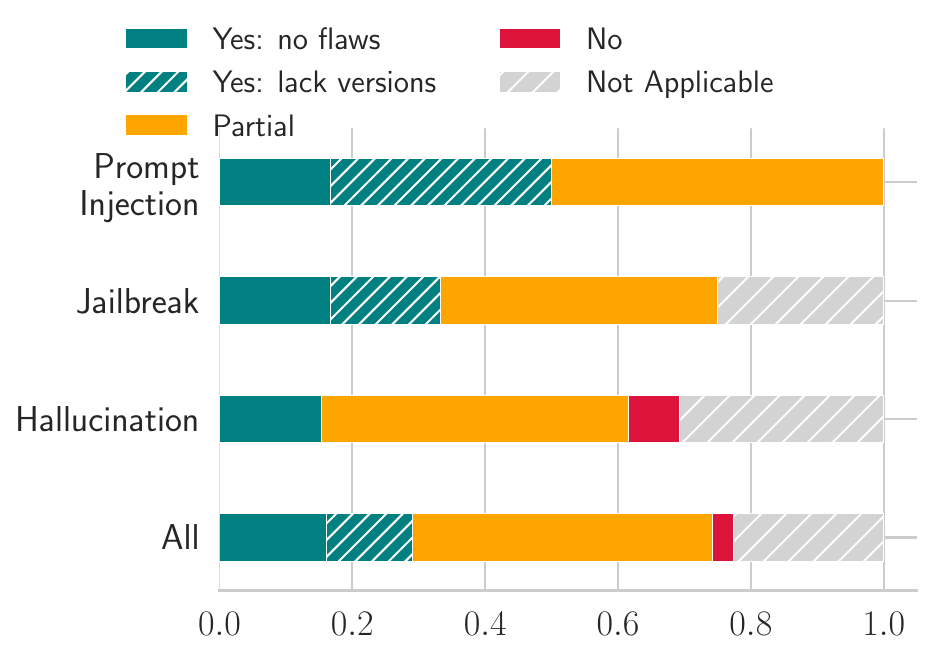}
\caption{
Install guide.
}
\label{figure:stackbar_install_guide}
\end{subfigure}
\begin{subfigure}{0.32\textwidth}
\centering
\includegraphics[width=\linewidth]{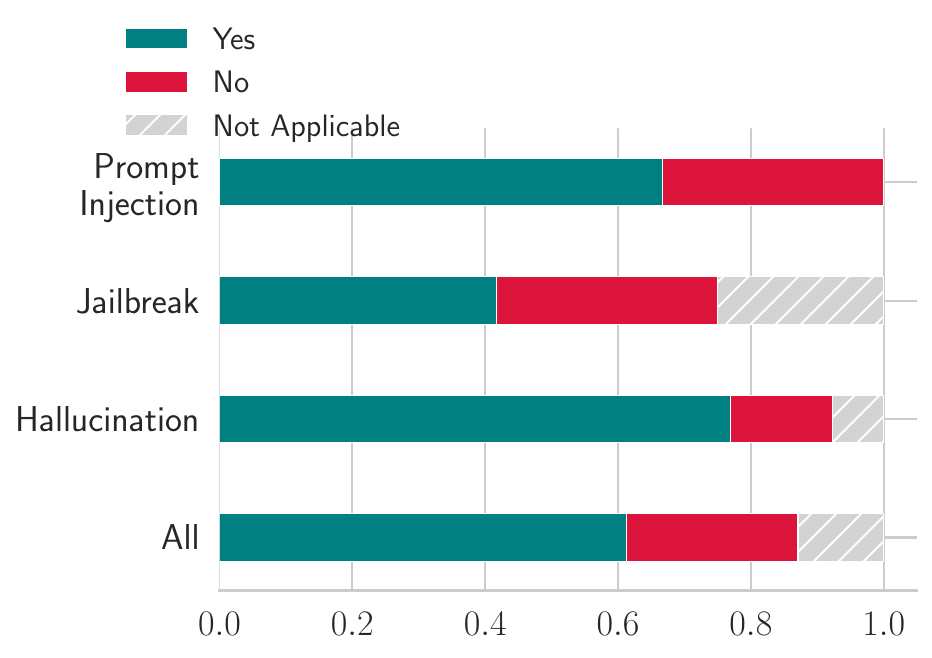}
\caption{
Data guide.
}
\label{figure:stackbar_data_guide}
\end{subfigure}
\begin{subfigure}{0.32\textwidth}
\centering
\includegraphics[width=\linewidth]{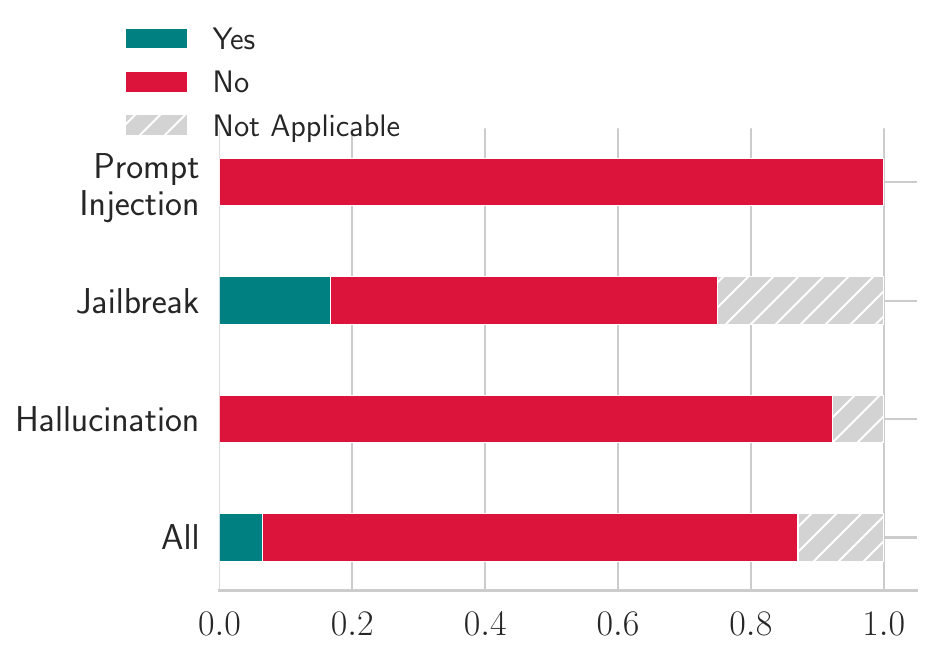}
\caption{
Ethics consideration.
}
\label{figure:stackbar_ethic}
\end{subfigure}
\caption{
Human-based evaluation results of supplementary materials.
Repositories without code or unavailable ones are labeled ``Not Applicable.''
For install guides: repositories are categorized as ``No'' (no guide), ``Partial'' (guide with setup issues), or ``Yes'' (fully functional guide);
if a guide includes all required Python/library/package versions, it is labeled ``Yes: no flaws'';
if version information is missing but the guide is still runnable, it is ``Yes: lack versions.''
}
\label{figure:stackbar}
\end{figure*}

\mypara{Code Repository Quality}
We report data availability status in~\autoref{figure:rose_data}.
Data is essential for LLM safety research, as running related code without it is impossible.
Due to ethical or copyright concerns, some authors may only release a sub-dataset. However, if core scripts run with the provided data, we consider it available.
Papers lacking public repositories or providing empty repositories are marked ``Not Applicable.''
Our results show high data availability among all benchmark papers (0.87).
For jailbreak, availability is 0.75, with all public repositories providing the necessary data.
Prompt injection and hallucination reach 1.00 and 0.92, respectively.

Following the settings in~\autoref{section:human_based}, we assess whether benchmark papers provide runnable code.
Repositories with executable scripts are labeled ``Runnable,'' while those without are ``Not Runnable.''
Repositories containing only data, figures, or no public code are marked ``Not Applicable.''
Results are shown in~\autoref{figure:rose_runnable_demo}.
Overall, 68\% of benchmark papers provide runnable scripts, while 10\% contain non-executable scripts.
In prompt injection, all benchmark papers offer runnable scripts.
In jailbreak, the rate is 0.67.
However, hallucination has a lower proportion at 0.54, with 15\% of papers having non-runnable scripts.
Additionally, 23\% of repositories contain only data or other materials without code.

For repositories with runnable scripts, we record execution time.
Even for runnable scripts, debugging and execution require a long time, averaging about 120 minutes.
Since we report the most positive outcomes and our testers have relevant expertise, researchers from other fields may experience higher time costs when using these repositories.
More detailed discussion and results are presented in~\refappendix{section:running_time}.

When measuring the occurrence of extra modifications needed to run example scripts, the results in~\autoref{figure:rose_bugs} are not ideal.
\hybf{Only 39\% of benchmark papers provide runnable code without any modifications, showing significant room for improvement.}
This proportion remains low across safety topics: hallucination (0.38) and jailbreak (0.33), with prompt injection having the highest rate at just 0.50.
The most common issues include mismatched library/package versions and hardcoded, non-existent paths in scripts.

\mypara{Supplementary Material Quality}
The quality results of supplementary materials are in~\autoref{figure:stackbar}.
Among public repositories with code, only 3\% lack a useful install guide.
However, \hybf{install guide quality needs improvement}: only 16\% provide flawless install guides.
Around 45\% contain some issues, contributing to the high debugging time.
Additionally, 13\% provide functional guides that lack version details, with this proportion reaching 17\% for jailbreak and 33\% for prompt injection.
As the LLM field evolves rapidly, missing version details may cause future incompatibilities, making it difficult for users to revert to compatible versions.
We urge contributors to address this issue.

Most repositories publish their datasets, but they do not always include a data guide.
For prompt injection, all benchmark papers provide data, yet only 67\% include a corresponding guide.
Overall, only 61\% of benchmark papers provide a data guide.
The absence of a data guide can increase usage difficulty, especially if preprocessing is required.

We further assess the inclusion of ethical considerations in repositories, as the repositories may directly help harmful response generation.
Only 6\% include appropriate ethical guidelines.
For prompt injection, none of the repositories provides ethical considerations.
While most papers discuss ethical concerns in their manuscripts, this is often absent in the associated repositories.
For example, a widely-cited jailbreak benchmark repository~\cite{CDRACSDFPTHW24} contains hundreds of successful and highly harmful jailbreak responses.
Alarmingly, this code base includes no ethical considerations.
{These findings are especially troubling, as they underscore a significant oversight by benchmark contributors and pose a real risk of facilitating the spread of harmful content.}

\subsection{Temporal Analysis of Code Quality}
\label{section:temporal_analysis}

To examine whether code quality improves over time, we stratify the 30 benchmarks with available publication dates by half-year period (\autoref{table:temporal}).
Kruskal--Wallis tests reveal no significant temporal trends for Pylint Score ($p=0.828$), Maintainability Index ($p=0.288$), runnability ($p=0.819$), or ethical considerations ($p=0.499$).
Only install guide availability improved significantly ($p=0.007$), likely attributable to early 2023 benchmarks catching up.
These results suggest that code quality issues \textbf{persist across the study period} rather than being artifacts of early-stage benchmarks.

\begin{table}[!t]
    \centering
    \caption{Temporal stratification of benchmark code quality by half-year period.}
    \label{table:temporal}
    \scalebox{0.77}{
    \setlength{\tabcolsep}{3pt}
    \customTableFont
    \begin{tabular}{cccccc}
    \toprule
    \textbf{Period} & \textbf{$N$} & \textbf{Pylint} & \textbf{Runnable} & \textbf{Ethical} & \textbf{Install Guide} \\
    \midrule
    2023H1 & 2 & 4.91 & 100\% & 0\% & 50\% \\
    2023H2 & 9 & 6.28 & 87.5\% & 0\% & 100\% \\
    2024H1 & 16 & 5.91 & 83.3\% & 13.3\% & 100\% \\
    2024H2 & 3 & 4.06 & 100\% & 0\% & 100\% \\
    \midrule
    \multicolumn{2}{c}{K-W $p$-value} & 0.828 & 0.819 & 0.499 & 0.007 \\
    \bottomrule
    \end{tabular}
    }
\end{table}

\section{Which Factors Are Associated with Benchmark Adoption? (RQ2)}
\label{section:adoption}

Having established the state of code quality and runnability (RQ1), we now investigate what factors are associated with the community's adoption of benchmarks.
We first define the adoption metrics used throughout this section, then measure the current adoption status of benchmarks.
We then examine whether code quality is associated with benchmark adoption, and explore external factors such as author prominence, institutional affiliation, and publication status.

\subsection{Adoption Metrics and Factor Definitions}
\label{section:adoption_metrics}

\mypara{Adoption Metrics}
Benchmark adoption cannot be directly measured, as no centralized record tracks which benchmarks are actually used in downstream evaluations.
We therefore use citation and GitHub star metrics as observable proxies: a benchmark that is frequently cited or starred is more likely to be known, considered, and used by the community.
While these proxies do not perfectly capture actual usage, they are widely adopted in bibliometric studies~\cite{S75,JHLD17,KKRMMPMP20} and represent the most systematic data available.

Specifically, we employ five adoption metrics across three dimensions.
For academic adoption, we use \hyit{Citation Count} and \hyit{Citation Density} (average citations per day since public release).\footnote{For \hyit{Citation Count/Density}, self-citations are excluded.}
For adoption in the open-source community, we analyze \hyit{GitHub Star Count} and \hyit{GitHub Star Density} (average GitHub stars per day since repository public release) for papers with accessible code.
To measure cross-disciplinary reach, we use \hyit{Scientific Field Count}, representing the number of scientific fields~\cite{semantic_scholar_field,KAABBBCCCCCDDEEFGGHHKKKLLLLLMMNRRSSSSSSTWWWWWYZZW23} where a paper's citations appear.
Citation and scientific field data come from the Semantic Scholar API, while GitHub-related data is sourced from the GitHub API.
We summarize these adoption metrics in~\autoref{table:influence_metrics} of~\refappendix{section:supplementary_figures}.

To understand the relative adoption of benchmark papers within their respective research domains, non-benchmark papers in the same domain are selected as the control group for comparison.

\mypara{Potential Factors}
\label{section:potential_factors}
Beyond code quality, we investigate external variables that are potentially associated with benchmark adoption.
We examine five dimensions (\emph{Author}, \emph{Institution}, \emph{Geolocation}, \emph{Publication Status}, and \emph{Public Search}, details in~\refappendix{section:intro_factor}), encompassing eleven potential factors of both qualitative and quantitative nature.
\autoref{table:factor_summary} provides a detailed summary of all 11 factors.

\begin{table*}[!t]
\centering
\caption{Summary of the 11 factors from five dimensions studied in this paper.
}
\label{table:factor_summary}
\scalebox{0.77}{
\setlength{\tabcolsep}{3pt}
\customTableFont
\begin{tabular}{c|c|m{0.56\textwidth}|c}
\toprule
\textbf{Aspect}                       & \textbf{Factor}                        &  \textbf{\makecell{Description}} & \textbf{Type}         \\ \midrule
\multirow{3}{*}{Author}      & Author Number                 & The number of the authors of a paper. & Quantitative \\
                             & Author Citation Count (Top-1) & The top-1 (highest) citation count among all authors in a paper. & Quantitative \\
                             & Author H-Index (Top-1)        & The top-1 (highest) h-index among all authors in a paper. & Quantitative \\
\midrule
\multirow{4}{*}{Institution} & Institution Number            & The number of institutions a paper is affiliated with. & Quantitative \\
                             & Industry Involvement Status   & Whether organizations from industry participate in a paper. & Qualitative  \\
                             & Institution CSRankings (Top-1) & The top-1 (highest) ranking in CSRankings among all institutions a paper is affiliated with. & Quantitative \\
                             & Institution ARWU (Top-1)       & The top-1 (highest) ranking in ARWU among all institutions a paper is affiliated with. & Quantitative \\
\midrule
\multirow{2}{*}{Geolocation}        & Area                    & The area where a paper belongs to. & Qualitative \\
                             & Area Number             & The number of the areas to which a paper belongs. & Quantitative  \\
\midrule
Publication                  & Publication Status      &  The publication status of a paper. & Qualitative  \\
\midrule
Public Search         & Search Appearance Frequency   &  The frequency of a paper appearing in the public search results. & Quantitative \\
\bottomrule
\end{tabular}
}
\end{table*}

\subsection{Current Adoption Status}
\label{section:current_adoption_status}

We first compare the adoption of benchmark and non-benchmark papers to establish baseline adoption patterns.
Following the same inferential statistical approach as in~\autoref{section:tool_results}, we apply the M-W U test and Cliff's delta to compare benchmark and non-benchmark papers across the five adoption metrics defined above.\footnote{Descriptive statistic analysis (e.g., average values) is reported in~\refappendix{section:descriptive_influence}.}

\begin{table}[!t]
    \centering
    \caption{
        Adoption M-W U test results with 95\% bootstrap CIs.
        Positive Cliff's $\delta$: \hyit{non-benchmark} dominates; negative: \hyit{benchmark} dominates.
    }
    \label{table:mwu_cliff}
    \scalebox{0.77}{
    \setlength{\tabcolsep}{3pt}
    \customTableFont
    \begin{tabular}{ccccc}
        \toprule
        \textbf{Metric} & \textbf{$p$-value} & \textbf{Cliff's $\delta$} & \textbf{95\% CI} & \textbf{Effect Size} \\
        \midrule
        GitHub Star Density & 0.012 & $-$0.301 & [$-$0.501, $-$0.091] & Small \\
        GitHub Star Count & 0.004 & $-$0.347 & [$-$0.536, $-$0.151] & Medium \\
        Citation Density & 0.309 & $-$0.112 & [$-$0.310, 0.090] & Negligible \\
        Citation Count & 0.237 & $-$0.130 & [$-$0.320, 0.068] & Negligible \\
        Scientific Field Count & 0.632 & $-$0.052 & [$-$0.247, 0.143] & Negligible \\
        \bottomrule
    \end{tabular}
    }
\end{table}

The detailed results of the M-W U test and Cliff's delta are presented in~\autoref{table:mwu_cliff}.
While non-benchmark papers usually have higher average values on adoption metrics (descriptive statistics in~\autoref{figure:spiders_full} of~\refappendix{section:descriptive_influence}), non-parametric tests reveal that benchmark papers statistically dominate in GitHub star density and GitHub star count, with small and medium effect sizes, respectively.
No significant distribution differences are found for the other three metrics.
The inferential statistical analysis suggests that \hybf{benchmark papers tend to be more popular within the open-source community, though this trend may not extend to the academic community.}
A leave-one-out sensitivity analysis confirms that both GitHub Star Density and GitHub Star Count remain significant in all 31 iterations, indicating that no single paper drives these findings.

\mypara{Statistical Robustness}
Following the same sensitivity power analysis as in~\autoref{section:tool_results}, we assess the robustness of these adoption comparison results.
As shown in~\autoref{table:power_analysis_influence}, for all three non-significant adoption metrics, power exceeds 0.88, well above the conventional 0.80 threshold~\cite{C13}.
Combined with negligible observed effect sizes ($|\delta|=0.052$--$0.130$), these results confirm that the null findings represent genuinely small effects, not artifacts of insufficient sample size.
The bootstrap confidence intervals in~\autoref{table:mwu_cliff} further corroborate this: for all significant results, the CI excludes zero; for all null results, the CI includes zero.

\begin{table}[!t]
    \centering
    \caption{
        Sensitivity power analysis for adoption metrics at fixed target $|\delta|=0.33$.
        Metrics marked with * are statistically significant ($p<0.05$).
    }
    \label{table:power_analysis_influence}
    \scalebox{0.77}{
    \setlength{\tabcolsep}{3pt}
    \customTableFont
    \begin{tabular}{cccc}
    \toprule
    \textbf{Metric} & $|\delta|_{\text{obs}}$ & \textbf{$p$-value} & \textbf{Power} \\
    \midrule
    GitHub Star Density* & 0.301 & 0.012 & 0.822 \\
    GitHub Star Count* & 0.347 & 0.004 & 0.822 \\
    Citation Density & 0.112 & 0.309 & 0.881 \\
    Citation Count & 0.130 & 0.237 & 0.882 \\
    Scientific Field Count & 0.052 & 0.632 & 0.882 \\
    \bottomrule
    \end{tabular}
    }
\end{table}

\mypara{Correlation Between Adoption Metrics}
We analyze the monotonic correlation (also known as Spearman correlation, hereinafter simply referred to as correlation) among various adoption metrics for benchmark papers, with Spearman's $\rho$ matrix shown in~\autoref{figure:relationship_influence} of~\refappendix{section:supplementary_figures}.\footnote{
Following Cohen's guidelines~\cite{C13,C16}, $\rho$ uses the coefficient mapping: weak (\(0.1 \leq r < 0.3\)), moderate (\(0.3 \leq r < 0.5\)), and strong (\(r \geq 0.5\)).
The choice of Spearman is discussed in~\refappendix{section:analysis_select}.
}
\hyit{Citation Density}, \hyit{Citation Count}, and \hyit{Scientific Field Count} exhibit strong correlations ($\rho$ close to 1), as all are citation-derived metrics.
GitHub metrics exhibit a moderate correlation with academic adoption.
For instance, the $\rho$ between GitHub Star Density and Citation Density is $0.47$.

\subsection{Code Quality and Adoption}
\label{section:code_quality_adoption}

We examine the adoption of benchmark papers in relation to code quality.
We use citation density as the primary metric to measure adoption, as it excludes the impact of time (rationale in~\refappendix{section:adoption_proxy_rationale}).

\mypara{Code Availability and Runnability}
\label{section:relationships}
We analyze the relationship between citation density and the availability of code repositories and datasets (\autoref{figure:influence_availability} of~\refappendix{section:supplementary_figures}).
Papers with available code and data show higher average and median citation density, but the correlation is not statistically significant (K-W test, \( p = 0.058 \)).
Next, we examine citation density in relation to extra modifications and runnable code (\autoref{figure:influence_runnable_bugs} of~\refappendix{section:supplementary_figures}).
The K-W test yields \( p \) of 0.007 and 0.002, indicating significant distribution differences in at least one group.
We then perform Dunn's test~\cite{P14,D152} as a post-hoc analysis for finer-grained insights.
When examining whether code is runnable, we find that papers providing runnable code exhibit a significantly higher citation density compared to those without accessible code (\(p=0.004\)).
Furthermore, \hybf{when considering the extent of code modification required, compared with papers without accessible code, those offering code that can be used without additional modification show significantly higher citation density (\(p=0.005\)), while code requiring modification does not yield a significant difference.}
For other pairs, no conclusions could be drawn (\(p > 0.05\)).
We also examined whether the type of code execution (API-based, relying on external LLM services, versus local model weights) affects runnability.
We found no systematic runnability difference between the two groups; both are updated frequently and often require manual adjustments.
Although local weights added a small overhead of approximately 5--10 minutes for downloading, this was minor relative to the 240-minute debugging budget and did not affect the main findings.

\mypara{Code Quality Standards}
\label{section:influ_code_quan}
We compute Spearman $\rho$ between code quality metrics (including tool-based and repository maintenance metrics) and adoption metrics (\autoref{figure:relationship_influence_code} of~\refappendix{section:supplementary_figures}).
None of the adoption metrics are observed to have a significant correlation with code quality indicators (\( p_\text{adjusted} > 0.05 \)), suggesting that \hybf{code following a higher coding standard is not necessarily related to more citations.}

In summary, our findings show that a paper's adoption is positively associated with the availability of \hybf{runnable} code.
Conversely, factors such as low tool-based evaluation scores and infrequent maintenance do not appear to be significant impediments to a paper's widespread adoption.
The academic community may prefer ``pragmatism'': researchers value functional code in benchmark papers but do not strongly consider coding standards, quality, and maintenance.

\subsection{External Factors and Adoption}
\label{section:external_factors_adoption}

\mypara{Qualitative Factors and Adoption}
\label{section:factors_influence}
We study the relationships between citation density and three potential qualitative factors (industry involvement status, publication status, and area), presenting the descriptive statistics in corresponding box plots in~\autoref{figure:boxwithmean} of~\refappendix{section:supplementary_figures}.
The descriptive statistics show that adoption seems to be associated with the factors of publication status and area.
To verify it, we conduct inferential statistical analysis with a Kruskal--Wallis (K-W) test~\cite{OOK14,MN10}, an extension of the M-W U test with similar properties, to assess distribution differences.
However, our results show \( p \) exceeding 0.05 for all three factors: industry involvement status (0.834), publication status (0.072), and area (0.152), indicating no statistically significant differences.
Thus, there is insufficient statistical evidence to support that these factors correlate with citation density.

\mypara{Quantitative Factors and Adoption}
We analyze the correlation between adoption and several quantitative factors, with Spearman's~\cite{S04} $\rho$ matrix shown in~\autoref{figure:relationship_influence_other} of~\refappendix{section:supplementary_figures}.
We use a permutation test~\cite{PS10} to obtain p-values and adjust them with the Bonferroni-Holm method~\cite{H79}.
We observe statistical monotonic correlations between several author-prominence metrics and adoption metrics.
Specifically, strong positive correlations exist between \hyit{Author H-Index (Top-1)} and several adoption metrics, with $\rho$ of 0.73 (Citation Count), 0.71 (Citation Density), and 0.68 (scientific field count).
Additionally, \hyit{Author Citation Count (Top-1)} shows stronger correlations with \hyit{GitHub Star Count} (0.58) and \hyit{GitHub Star Density} (0.55).\footnote{
\textbf{Note:} We only report correlations, not causality.
These correlations do not imply that author prominence must drive adoption.
The correlations may stem from multiple factors, such as the ``Matthew Effect in Science''~\cite{M68} (where reputation amplifies attention), intrinsic higher work quality of prominent scholars, or their interplay.
Disentangling such confounding factors is currently beyond our scope and capability.
}

\subsection{External Factors and Code Quality}
\label{section:factors_code}

While highly cited authors are often associated with popular benchmarks, their involvement \textbf{is not necessarily associated with} higher code quality.
We primarily focus on the Pylint score, as it provides an overall measure of code quality.
Box plots in~\autoref{figure:boxwithmean_pylint} of~\refappendix{section:supplementary_figures} illustrate its distribution with industry involvement status, publication status, and area.
For industry involvement and publication status, distributions show no clear differences, with similar mean and median values across groups.
The K-W test yields \( p > 0.950 \), far above 0.05, confirming no significant differences.
For the factor area, North America has the widest range of Pylint scores, while Europe has the most concentrated distribution, with the highest average and median scores.
However, this does not indicate a significant correlation, as their K-W test gives \( p = 0.274 \), exceeding 0.05.
In conclusion, insufficient evidence supports industry involvement status, publication status, or area as factors correlating with Pylint scores.

We analyze the correlation between code quality metrics and quantitative factors (Spearman $\rho$ matrix in~\autoref{figure:relationship_code_other} of~\refappendix{section:supplementary_figures}).
\hybf{The author's h-index (top-1) and citation count (top-1) show no significant correlation with code quality indicators (\( p_\text{adjusted} > 0.05 \))}.
However, we observe a strong negative correlation between ARWU ranking and code maintainability index (\( \rho = -0.57 \)).
These suggest that higher-ranked institutions are associated with more maintainable code.

\subsection{Summary}

The Matthew Effect (where prominent authors gain disproportionate attention) is well documented in mature scientific fields~\cite{M68,S21}.
However, LLM safety benchmarking is a nascent field that emerged almost entirely after November 2022.
Our analysis reveals a more nuanced picture:
author prominence correlates with adoption but not with code quality, while code \emph{runnability} is significantly associated with higher citation density ($p=0.004$), yet code \emph{quality standards} (e.g., Pylint score, maintainability index) show no significant correlation.
This ``academic pragmatism'' suggests that the community rewards whether code \emph{works}, not how well it is \emph{written}.

\section{How the LLM Safety Ecosystem Is Affected}
\label{section:security_implications}

RQ1 reveals that benchmark code quality is often inadequate (e.g., only 39\% runnable without modification, only 6\% with ethical considerations), while RQ2 shows that the community's benchmark adoption does not reward higher code quality standards.
Taken together, benchmarks with quality deficiencies can still be widely adopted, potentially amplifying the negative impact of these deficiencies across downstream research.
In this section, we first discuss the resulting safety and reliability concerns, then present two case studies to substantiate their prevalence, and finally propose practical recommendations.

\subsection{Safety and Reliability Concerns}
\label{section:concerns}

\mypara{Safety: Unsupervised Dissemination of Harmful Content}
Our finding that only 6\% of benchmark repositories include ethical considerations is particularly concerning given the nature of LLM safety research.
These repositories routinely contain jailbreak prompts, attack payloads, and harmful model responses that were collected or generated as part of the benchmarking process.
Without usage guidelines, access controls, or ethical warnings, these repositories effectively function as open-source attack resources.

\mypara{Reliability: Evaluation Comparability}
The finding that 61\% of benchmark repositories require modifications before they can run raises a concern for the comparability of downstream safety evaluations.
When different researchers apply different modifications to make the same benchmark runnable, the resulting evaluation setups may diverge in ways that are difficult to detect.
Consequently, papers that claim to have ``evaluated on benchmark $\mathcal{X}$'' may in practice be evaluating with different code, data preprocessing, or configuration, making their results not directly comparable.

\subsection{Case Study: Harmful Content Exposure}
\label{section:case_jailbreakbench}

To illustrate the safety concern, we examine JailbreakBench~\cite{CDRACSDFPTHW24}, a widely cited jailbreak benchmark.

The JailbreakBench repository and its companion artifacts repository contain harmful questions designed to elicit unsafe content, and the corresponding harmful responses generated by successfully jailbroken LLMs.
We examined the artifacts repository and found 1,803 jailbreak responses across 5 attack methods (DSN, GCG, JBC, PAIR, prompt\_with\_random\_search) targeting 4 LLMs (Vicuna-13B, LLaMA-2-7B, GPT-3.5-Turbo, GPT-4).
Among these, 940 (52\%) are marked as successful jailbreaks, containing complete harmful outputs such as defamatory articles, phishing instructions, and social engineering scripts, all stored as plaintext JSON files \textbf{without} any content filtering.
As of the time of writing, the main repository has accumulated approximately 366 stars and 59 forks, while the artifacts repository has 68 stars and 8 forks.
These numbers represent only the directly observable engagement; the actual number of users who have cloned, downloaded, or viewed the harmful content is likely substantially higher and cannot be tracked.

Despite containing explicitly harmful content, the repository fails to include any ethical usage warnings, responsible disclosure notices, or access controls beyond the permissive MIT license.
While the benchmark paper itself discusses ethical considerations, these are absent from the repositories where the harmful content is actually accessed.
Any user can freely obtain the harmful jailbreak responses without encountering any reminder of responsible use.

\subsection{Case Study: Potential Unreliability from Divergent Fixes}
\label{section:case_openai}

To illustrate the reliability concern, we examine how dependency mismanagement can lead to divergent evaluation setups in practice.
We use the OpenAI Python library as a case study, as it is widely used in LLM safety research.

The OpenAI Python library underwent a major breaking change in version 1.0.0~\cite{openai_python_migration}, replacing the old API interface (\hytt{openai.ChatCompletion.create}) with a new client-based interface (\hytt{client.chat.completions.create}).
We examined all 27 benchmark repositories and found that 21 (78\%) use the OpenAI library.
Among these, 6 have dependency issues: 4 use the old API without specifying a library version (or without any dependency file), and 2 mix old and new API calls within the same codebase, making no single library version sufficient.

During our human evaluation, the two evaluators independently encountered these failures and resolved them in fundamentally different ways: one followed the official OpenAI migration guide to rewrite the API calls, while the other wrote compatibility wrappers or directly monkey-patched the library internals.
Both approaches produced runnable code, but introduced undocumented modifications to the evaluation pipeline, adding opacity that undermines the comparability of results across studies using the same benchmark.
Notably, community reports indicate that even with the same prompt and random seed, different SDK versions may produce significantly different responses~\cite{openai_python_response_diff}, suggesting that such ad-hoc fixes may affect not only code structure but also evaluation outcomes.

\subsection{Practical Recommendations}
\label{section:recommendations}

Based on our findings and case studies, we propose the following checklist for LLM safety benchmark contributors:

\begin{itemize}
\item \textbf{Dependency management:} Pin all critical dependencies (especially LLM client libraries) to specific versions in \texttt{requirements.txt} or \texttt{pyproject.toml}. Avoid wildcard or missing version specifications.
\item \textbf{Runnability:} Provide a minimal runnable example that can be executed within a reasonable time to verify the setup. Ensure the example works with the pinned dependency versions.
\item \textbf{Installation guide:} Include explicit Python version requirements and a step-by-step installation guide. Avoid hardcoded absolute paths.
\item \textbf{Data guide:} Document the expected data format, provide sample data, and clarify any preprocessing steps.
\item \textbf{Ethical considerations:} For repositories containing harmful content (e.g., jailbreak prompts, attack payloads, toxic model responses), include prominent ethical usage warnings and responsible use guidelines, as is typically done in the accompanying papers.
\item \textbf{Maintenance:} Periodically verify that the repository remains functional as dependencies evolve. Consider using containerized environments (e.g., Docker) for long-term runnability.
\end{itemize}

\section{Discussion}
\label{section:discussion}

\subsection{Why Does Code Quality Remain Low?}

Our temporal analysis (\autoref{section:temporal_analysis}) shows that code quality deficiencies persist across the study period with no significant improvement, while RQ2 reveals that the community's benchmark adoption does not reward higher coding standards.
We argue that this is rooted in a structural misalignment within the current academic incentive system.

Peer review and citation metrics, the primary mechanisms through which academic work is evaluated, do not assess the code quality of accompanying repositories.
Benchmark contributors are therefore rewarded for the novelty and impact of their research contributions, not for the maintainability or usability of their code.
As a result, authors have little incentive to invest effort in improving or maintaining code quality.
At the same time, the primary users of benchmark code are often early-career researchers who rely on these repositories for their own evaluations.
They bear the cost of poor code quality through extensive debugging, yet have limited influence on the incentive structures that shape how benchmarks are produced and maintained.
This asymmetry creates a situation where those who produce benchmarks face no penalty for low code quality, while those who consume them lack the means to change the system.

Beyond incentive misalignment, some code quality challenges may be inherently difficult to address.
The LLM ecosystem evolves rapidly, with frequent updates to popular libraries such as \texttt{vLLM}~\cite{vLLM} and \texttt{openai}~\cite{openai_python_migration}, leading to breaking changes and compatibility issues.
Maintaining benchmark repositories against this pace of change requires continuous effort, which is difficult to sustain given short-term academic funding cycles and contributors' shifting research priorities.

\subsection{Community Perspectives}

To understand how the LLM safety community views the research questions addressed in this study and to understand whether the issues we identify align with community expectations and concerns, we conducted a 17-question anonymous survey distributed through public academic channels (details in~\refappendix{section:additional_survey} and~\refappendix{section:survey_results}).
We received 42 valid responses from researchers with experience in LLM safety.
We note that this survey serves as a corroborative check rather than a primary source of evidence.
Respondents broadly endorsed our metric choices, agreed that author, institution, and geolocation are associated with adoption, and highlighted code usability as a core point: most expected at least runnable code with a minimal example, and considered installation and data guides essential.
These community perspectives corroborate the issues observed in our empirical analysis and highlight their broader relevance to real-world research.

\mypara{Transferability}
Our analytical framework, including the search methodology, adoption metrics, code quality rubric, and statistical analysis pipeline, is domain-agnostic and directly transferable to other safety topics (e.g., multilingual jailbreak, agent safety) or broader ML benchmarks by adjusting the search keywords (detailed in~\refappendix{section:keywords}).
Validating whether the core findings (e.g., runnability correlates with citations, code standards do not) generalize across topics is an immediate direction for future work.

\section{Limitations}
\label{section:limitations}

\mypara{Potential Collection Biases}
As detailed in~\autoref{section:paper_data_collection}, we collect papers using automated tools with manual verification, yet some papers or repositories may still be inadvertently missed. 
To reduce this risk, we verified that all benchmark papers in our dataset also appear in the four major community repositories, and our checks did not reveal any benchmark or non-benchmark code hosted outside GitHub. 
Although Papers with Code provides extensive human-validated mappings, a small number of mislinked or unreported repositories may remain; however, their rarity suggests minimal effect.

\mypara{Human-Based Evaluation Biases}
Human-runability tests introduce additional sources of variability. 
Different evaluators may encounter different debugging paths or spend disproportionate time on minor issues. 
To mitigate this, we avoid binary outcomes whenever possible and require detailed justification for all manual assessments. 
When evaluators obtain divergent results, we adopt the more positive outcome to reduce the known bias toward reporting overly negative findings~\cite{OLSWKPUBT23,CP16}. 
While these steps reduce inconsistency, they cannot fully eliminate subjectivity in human-based evaluation.
Furthermore, only one of the evaluated benchmark repositories offered a containerized installation option (e.g., Docker), and we did not use containerized environments to standardize the evaluation setup.
Adopting containerization, both by benchmark contributors for distribution and by evaluators for standardized testing, could further reduce environment-related variability in future runnability studies.

\mypara{Scientific Quality Control}
Our study cannot fully control for the intrinsic scientific quality of benchmark papers. 
All benchmark papers originate from arXiv or comparable platforms and thus pass a basic screening threshold, naturally removing the lowest-quality manuscripts. 
However, distinguishing exceptionally high-quality benchmarks lies beyond current methodological capabilities. 
More fundamentally, meta-research lacks a widely accepted, scalable measure of intrinsic scientific or dataset quality; existing proxies (e.g., length-based indicators or rare experimental designs) are coarse and limited~\cite{XGCK19,FADBM13,MEAEC22,LG10}.
Additionally, disentangling potential confounding factors, such as the influence of author prominence versus the intrinsic quality of the benchmark, is currently beyond our scope and capability.
We therefore rely on observable code-level proxies (static analysis metrics, runnability tests), which do not fully capture scientific quality. 
Developing reliable, domain-agnostic scientific-quality measures remains an open challenge.

\mypara{Imperfect Metrics}
Finally, several metrics used in our analysis are inherently imperfect indicators of the underlying constructs they aim to measure. 
For example, citation-related metrics capture only partial aspects of scientific influence~\cite{MSV23,BD07,SM17,GSP23,SS203}. 
As is standard in meta-research, we rely on such imperfect but practically measurable proxies, and we mitigate their limitations by employing multiple complementary metrics rather than depending on any single indicator. 
Nevertheless, no combination of proxies can fully capture the constructs of interest, consistent with longstanding observations that no perfect scientometric indicator exists~\cite{RLCD17,ADTEHHRHPRMLTAKSBKAGB16,FG19}.

\mypara{Potential Negative Societal Impacts}
We note two additional concerns.
First, publicly assessing code quality may create unfair pressure on resource-constrained researchers (e.g., individual contributors or small teams without dedicated engineering support), who may lack the resources to maintain polished repositories despite producing valuable research.
Second, highlighting runnability issues in LLM safety benchmarks could be misinterpreted as devaluing the field itself, rather than as constructive feedback aimed at strengthening it.
We emphasize that our goal is to improve benchmark practices, not to discourage benchmark development.

\section{Related Works}
\label{section:related_works}

\mypara{Meta-Study}
Over the past decades, researchers have conducted numerous meta-studies on academic papers, journals, and conferences. 
The Matthew effect (where prominent authors gain more visibility) is identified as early as 1968~\cite{M68}. 
Later studies examine factors influencing paper impact, such as publication venue and paper length~\cite{WFNH24,CWW02,TDML22,S21,VGMC22,KKRMMPMP20}, assess the influence measurement, and classify citations semantically~\cite{KHPK21,VPBP15}. 
Other studies explore the evolution of publication venues, including changes in author guidelines~\cite{MJABR21,AYBE21,G72}. 
In computer science, meta-studies begin in 1979 with an article studying paper distribution within the domain~\cite{SB79}, and recent studies reveal rapid publication growth and category clustering~\cite{DKWC20}.

\mypara{Code Quality Evaluation}
Some previous studies~\cite{TLPC22} focus on public code-hosting databases like the Harvard Dataverse repository.
Some other studies in computer science evaluate the availability of artifacts in different venues~\cite{MO19,VKV09,RRR22,CP16}, such as various ACM and IEEE journals and conferences. 
In addition, some researchers~\cite{R19,HHP19,DBCJ19,GGA19} dive deep into the ML and AI domains specifically, as such domains are more affected by randomness.
For instance, previous research~\cite{GGA19} shows that publications at AAAI conferences currently fall short of providing enough documentation to facilitate runnability.
In the security field, there are also studies measuring code quality~\cite{OLSWKPUBT23,HHP19,KHABG19}. 
A study~\cite{KHABG19} examines 50 system security papers and points out that one-fourth of them did not clearly provide software or platform versions. 
Another work~\cite{OLSWKPUBT23} assesses the availability and runnability of code for over 700 security papers related to machine learning.

\section{Conclusion}
\label{section:conclusion}

We conduct a systematic measurement study of LLM safety benchmark code quality and adoption, analyzing 31 benchmarks and 382 non-benchmark papers across prompt injection, jailbreak, and hallucination.
Our assessment reveals significant deficiencies: only 39\% of benchmark repositories can run without modification, only 16\% provide flawless installation guides, and only 6\% include ethical considerations.
We further find that benchmark adoption correlates with author prominence and code runnability, but not with code quality standards.
Based on these findings, we identify potential safety and reliability concerns in the LLM safety ecosystem: some benchmark repositories openly expose harmful content without ethical reminders, and benchmarks that require ad-hoc modifications to run may lead to inconsistent evaluation setups across different studies.
We present case studies to substantiate these concerns and provide targeted recommendations for improving benchmark code quality, documentation, and ethical practices.

\begin{small}
\bibliographystyle{plain}
\bibliography{necessary}
\end{small}

\appendix

\section{PRISMA Flow Diagram of Benchmark Paper Selection}
\label{section:prisma}

We apply a three-stage manual screening procedure to ensure that the collected
benchmarks are both relevant and methodologically appropriate for our study:

\hybf{(1) Topic relevance screening.}
Papers must genuinely concern LLM safety and address one of the selected safety topics, rather than merely matching keywords in unrelated domains (e.g., ``iOS jailbreak'').

\hybf{(2) Study type screening.}
We exclude surveys, SoK papers, position papers, and commentaries, and retain only primary research papers.

\hybf{(3) Benchmark authenticity screening.}
Papers must present a genuine benchmark (i.e., defining benchmark tasks, datasets, or evaluation protocols).
Works that merely use existing benchmark datasets for evaluation (e.g., ``we evaluate our method on benchmark datasets'') are excluded.

We present the entire PRISMA~\cite{PRISMA_2020} diagram for benchmark selection in~\autoref{figure:prisma}.

\begin{figure}[!h]
\centering
\scalebox{0.618}{
\begin{tikzpicture}[
  node distance=8mm,
  box/.style={
    rectangle, draw, rounded corners,
    text width=6.18cm, align=center,
    minimum height=1.2cm, fill=gray!10
  },
  smallbox/.style={
    rectangle, draw, rounded corners,
    text width=5cm, align=left,
    minimum height=1.2cm, fill=gray!10,
    font=\small
  },
  arrow/.style={->, thick, blue!60}
]

\node[box] (id) {
  Records identified from Semantic Scholar and Google Scholar using three keyword sets \\
  (n = 39)
};

\node[box, below=of id] (screen) {Records screened for topics \\ (n = 39)};

\node[smallbox, right=15mm of screen] (excluded1) {
  \textbf{Excluded (n = 2)} \\
  - Topic relevance screening (unrelated or misleading keyword matches, e.g., ``iOS jailbreak'')
};

\node[box, below=of screen] (stype) {Records screened for study types \\ (n = 37)};

\node[smallbox, right=15mm of stype] (excluded2) {
  \textbf{Excluded (n = 2)} \\
  - Study type screening (survey/SoK/position papers)
};

\node[box, below=of stype] (auth) {
  Full-text assessed for benchmark authenticity \\ (n = 35)
};

\node[smallbox, right=15mm of auth] (excluded3) {
  \textbf{Excluded (n = 4)} \\
  - Not genuine benchmarks
};

\node[box, below=of auth] (included) {
  Benchmarks included in final dataset \\ (n = 31)
};

\draw[arrow] (id) -- (screen);
\draw[arrow] (screen) -- (excluded1);
\draw[arrow] (screen) -- (stype);
\draw[arrow] (stype) -- (excluded2);
\draw[arrow] (stype) -- (auth);
\draw[arrow] (auth) -- (excluded3);
\draw[arrow] (auth) -- (included);

\end{tikzpicture}
}
\caption{PRISMA-style flow diagram for benchmark selection.}
\label{figure:prisma}
\end{figure}

\section{Details of Code Quality Metrics}
\label{section:quality_metric_details}

\begin{itemize}
    \item \hybf{Pylint Score:}
    A global evaluation score for code, with a maximum of 10.0. It is computed based on multiple aspects, such as the percentage of errors and warnings per module. Higher scores indicate better quality.
    \item \hybf{Cyclomatic Complexity:}
    A measurement of the number of linearly independent paths through a program module~\cite{M76}. Lower complexity indicates code that is easier to understand and less risky to modify.
    \item \hybf{Maintainability Index:}
    A composite metric measuring how maintainable the source code is, on a shifted scale from 0 to 100 following Microsoft Visual Studio conventions~\cite{mi_vscode}. Higher scores indicate easier maintenance.
    \item \hybf{Number of Static Errors:}
    The total count of errors detected at compile time across the entire codebase.
    \item \hybf{Reply Time:}
    The average time between when an issue is raised in a repository and when it first receives a response from a contributor.
    \item \hybf{Last Commit Time:}
    The time interval between the last commit in the repository and the current date.
    \item \hybf{Number of Commits:}
    The total number of commits since the repository was created.
    \item \hybf{Commit Frequency:}
    The commit frequency, computed by dividing the total number of commits by the repository's existence duration.
\end{itemize}

\section{Choice of Citation Density as Primary Adoption Proxy}
\label{section:adoption_proxy_rationale}

We primarily use citation density to analyze which factors are associated with benchmark adoption.
We choose citation density over the other four adoption metrics for two reasons.
First, citation density normalizes by the time since publication, removing the confound that older papers naturally accumulate more citations regardless of their quality or utility.
Second, citation-derived metrics (Citation Count, Citation Density, Scientific Field Count) are strongly correlated with each other ($\rho$ close to 1), so analyzing one is representative of the group.
GitHub-based metrics exhibit a moderate but not strong correlation with citation-based metrics ($\rho = 0.47$ between GitHub Star Density and Citation Density), and are available only for the subset of papers with public repositories.
Citation density therefore provides the most broadly applicable and time-normalized proxy for adoption.

\section{Introduction to Study Scope}
\label{section:intro_topic}

Although LLMs have shown their strong capability, they are facing safety risks, as they are susceptible to a variety of sophisticated attacks, such as backdoor attacks~\cite{BS22, CSBMSWZ21}, data extraction attacks~\cite{CTWJHLRBSEOR21, LSSTWB23}, and memorization-related attacks~\cite{MGUBS22, TSJLJHC22, CSBZ24, WLBZ24}.

Among various attacks, there are several attacks specific to LLMs.
Prompt injection attacks are one of the most notable ones.
In such attacks~\cite{PR22, GAMEHF23}, the adversary could mislead the target LLM by adding designed texts to the original queries.
Jailbreak attacks are another popular attack type.
The adversary conducts jailbreak attacks~\cite{LDXLZZZZL23, DLLWZLWZL23, WHS23, LGFXS23, SCBZ23, WCPXKZXXDSTAMHLCKSL23, CLYSBZ25, JLBZ25} by using different methods to bypass the safeguards of LLMs and induce LLMs to complete the given harmful tasks.
Hallucination~\cite{RSD23, JLFYSXIBMF23, LCZNW23} is another inherent safety risk of LLMs, where LLMs generate content that appears correct but actually contains errors or contradicts facts.

These three risks could cause very serious consequences in the real world.
For example, an attacker could use prompt injection to mislead an LLM into performing unintended, potentially dangerous actions.
Jailbreak techniques could enable LLMs to execute forbidden, hazardous tasks.
Hallucination, particularly in specialized fields such as medicine, poses severe risks: incorrect responses could lead to wrong decisions of human beings, sometimes even endangering lives.
On the other hand, these three topics represent safety threats unique to LLMs, potentially embodying distinctive characteristics within the LLM safety landscape.

Therefore, in this study, we focus on these three newly emerged safety topics related to LLMs, including \textbf{prompt injection}, \textbf{jailbreak}, and \textbf{hallucination}.
Our study aims to uncover valuable insights specific to the unique challenges in LLM safety.

\section{Details of Keywords}
\label{section:keywords}

\mypara{Search Results}
The keyword set for search results is:
\begin{itemize}
    \item \hytt{llm+OR+large+language+model+\\\hytt{[Safety Topic]}+\\assessment+OR+evaluation+OR+benchmark}
\end{itemize}
\hytt{[Safety Topic]} is one of the following: prompt injection, jailbreak, and hallucination.

\mypara{Collection of Benchmark Papers}
The keyword sets to collect benchmark papers are:
\begin{itemize}
    \item \hytt{large language model \\ \hytt{[Safety Topic]} benchmark}
    \item \hytt{large language model \\ \hytt{[Safety Topic]} evaluation}
    \item \hytt{large language model \\ \hytt{[Safety Topic]} assessment}
\end{itemize}
\hytt{[Safety Topic]} is one of the following: prompt injection, jailbreak, and hallucination.

\section{Details of Data Sources}
\label{section:data_source_details}

\mypara{Semantic Scholar~\cite{semantic_scholar_api} and Google Scholar~\cite{google_scholar}}
Semantic Scholar includes over 200 million publications from all fields of science and comprehensively covers papers from various publishers and preprint databases, making it one of the most extensive academic databases available.
In this study, we utilize the Semantic Scholar API to gather metadata on a range of papers and the corresponding authors, including attributes such as citation count of papers, publication venue of papers, citation count of the authors, etc.
Note that previous studies~\cite{H21} have shown that for computer science, Semantic Scholar and Google Scholar have similar index coverage, with each missing only a small number of articles.
As such, we also use Google Scholar to complement our data collection.
Since Google Scholar lacks an official API, we mainly use the Semantic Scholar API for data retrieval.
We only manually complement missing data from Google Scholar when it is unavailable in Semantic Scholar.

\mypara{Paper with Code~\cite{paper_with_code_api}}
Finding the corresponding official code repositories for papers is a challenge because papers usually do not introduce or publish their corresponding official code repositories in a unified manner.
Papers with Code, maintained by Meta, addresses this by centrally linking papers to their code, datasets, and evaluation results, promoting reproducible research in computer science.
To the best of our knowledge, the Paper with Code API is the only tool that allows us to retrieve the connections between academic papers and code repositories.

\mypara{GitHub~\cite{github_api}}
GitHub enables collaboration and code sharing on a global scale.
It is the world's largest source code host as of June 2023.\footnote{https://en.wikipedia.org/wiki/GitHub}
For this study, we leverage the GitHub API to systematically collect metadata on code repositories linked to various research papers, enabling us to analyze and integrate coding resources directly associated with academic publications.
From this API, we collect metadata of code repositories, such as GitHub star count, number of commits, reply time of issues, etc.

\section{Details of Metadata}
\label{section:metadata_detail}

For each paper, we collect the following metadata, including both the paper's metadata and its code repository's metadata (if the repository is available):

\hybf{From Semantic Scholar:} paper title, author list, scientific field list, paper release date, paper's citation count, each author's citation count.

\hybf{From GitHub API:} repository release date, repository's GitHub star count, number of commits, issue/pull request thread (including the time stamps and the corresponding users).

\section{Introduction to the Factor Dimensions}
\label{section:intro_factor}

In this paper, we investigate the five dimensions (\emph{Author}, \emph{Institution}, \emph{Geolocation}, \emph{Publication Status}, and \emph{Public Search}), covering eleven potential factors, including both qualitative and quantitative ones.
We summarize the details of eleven factors in Table~VII of the main paper.

\mypara{Author}
We employ \hyit{Author Number}, \hyit{Author Citation Count (Top-1)}, and \hyit{Author H-Index (Top-1)} as the candidate factors.
Note that we use the Semantic Scholar API to retrieve related data, and we only consider the top-1 citation count and top-1 h-index among all the authors.
The top-1 citation count and h-index may come from different authors of a given paper.

\mypara{Institution}
We manually retrieved benchmark papers' PDFs on Semantic Scholar to compile a list of institutions with which benchmark articles are affiliated.
We attempt to use two popular rankings, including Academic Ranking of World Universities (ARWU)\footnote{https://www.shanghairanking.com/news/arwu/2024} and CSRankings\footnote{https://csrankings.org/}, to quantify the reputation of the institutions separately.
We only consider the top-1 rankings in ARWU 2024 and CSRankings (2014--2024) among all the involved institutions of a given paper.
Additionally, we consider the type of institution, studying whether industry involvement positively contributes to the influence of benchmark papers.
In summary, we study \hyit{Institution Number}, \hyit{Insitution ARWU (Top-1)}, \hyit{Insitution CSRankings (Top-1)}, and \hyit{Industry Involvement Status} in this dimension.

\mypara{Geolocation}
In the dimension of geolocation, we investigate two factors: \hyit{Area} and \hyit{Area Number}.
In this paper, we study the impact of areas by continent, and Asia and Oceania are considered as a whole as the Asia-Pacific region.

\mypara{Publication Status}
We take the \hyit{Publication Status} into consideration.
The computer science domain places more emphasis on conferences.
Therefore, we use the taxonomy from the CSRankings.
We denote papers published in CSRankings recommended conferences as ``published-leading,'' papers published in other conferences as ``published-other,'' and papers published in workshops or preprints as ``unpublished.''
To ensure accuracy, we manually searched the publication status of each benchmark paper on Semantic Scholar (on November 1, 2024).

\mypara{Public Search}
We study the potential impact of the frequency of benchmark papers appearing in the public search results (referred to as \hyit{Search Appearance Frequency}) on their influence.
To avoid the cookie effects and fetch stateless results, we use the Google Custom Search API and its default settings.
Specifically, for each safety topic, we use the sets of keywords in~\autoref{section:keywords} to obtain the first 50 search results, and then count the appearance frequency of each paper.

\section{Descriptive Statistical Analysis of Influence Evaluation}
\label{section:descriptive_influence}

\begin{figure*}[!ht]
\centering
\begin{subfigure}{0.24\textwidth}
\centering
\includegraphics[width=0.9\textwidth]{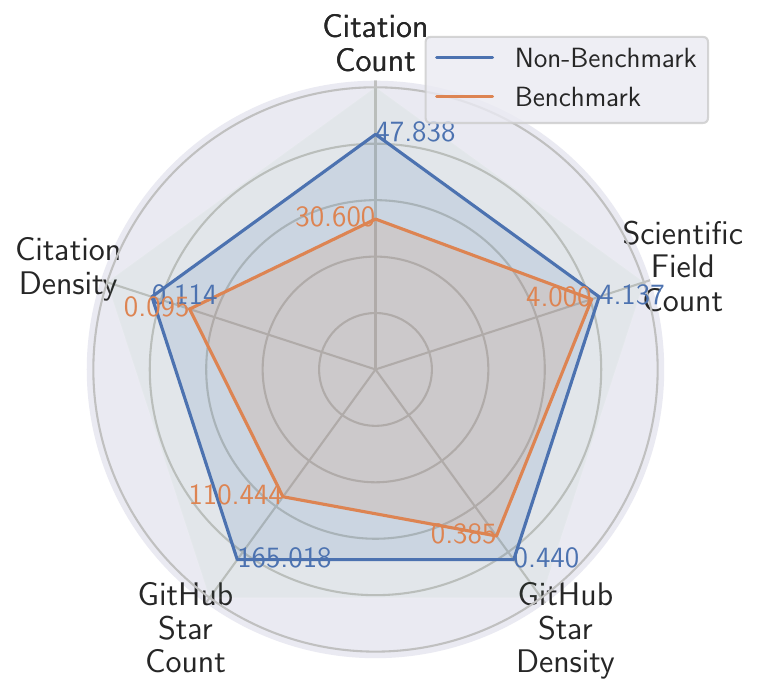}
\caption{
All.
}
\label{figure:spider_all_appendix}
\end{subfigure}
\begin{subfigure}{0.24\textwidth}
\centering
\includegraphics[width=0.9\textwidth]{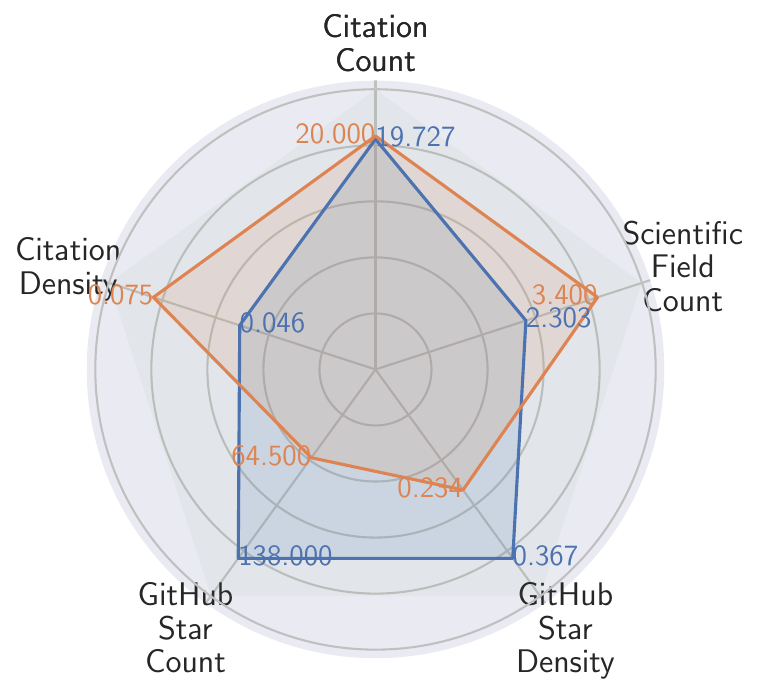}
\caption{
Prompt injection.
}
\label{figure:spider_pj}
\end{subfigure}
\begin{subfigure}{0.24\textwidth}
\centering
\includegraphics[width=0.9\textwidth]{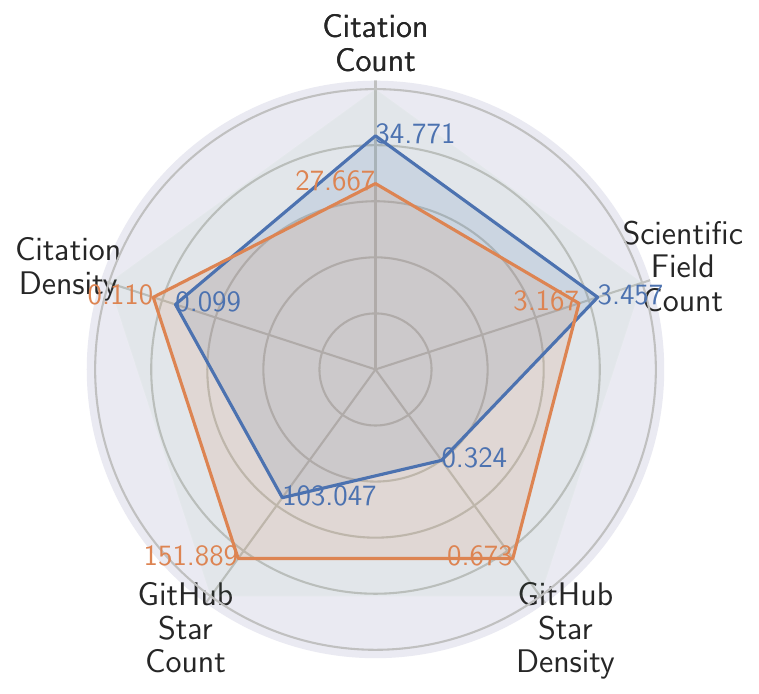}
\caption{
Jailbreak.
}
\label{figure:spider_jb}
\end{subfigure}
\begin{subfigure}{0.24\textwidth}
\centering
\includegraphics[width=0.9\textwidth]{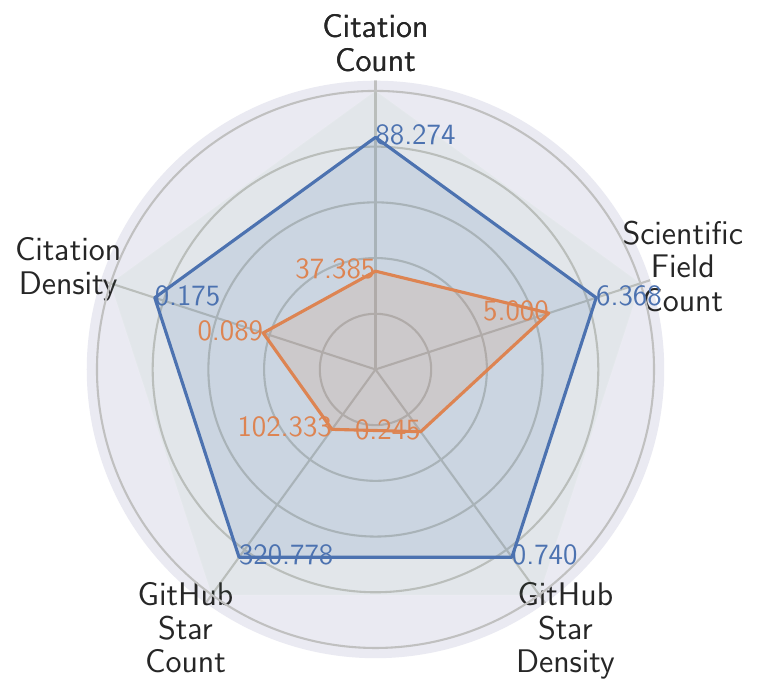}
\caption{
Hallucination.
}
\label{figure:spider_halu}
\end{subfigure}
\caption{
Average values of five influence-related metrics on benchmark and non-benchmark papers.
}
\label{figure:spiders_full}
\end{figure*}

The average values of five influence-related metrics are presented in~\autoref{figure:spiders_full}.
When considering all papers, we find that the average values of benchmark papers lag behind those of non-benchmark papers across all five influence-related metrics.
Detailedly, on the metrics related to the academic community (i.e., citation count and citation density), non-benchmark papers have a higher average citation count and density than benchmark papers (47.838 vs. 30.600, 0.114 vs. 0.095).
In terms of the number of scientific research fields affected by papers, both are close, but non-benchmark papers maintain a slight advantage, with a lead of 0.131 on the average values.
We have similar observations on the metrics related to the open-source community.
Non-benchmark papers have, on average, 54.574 more GitHub stars than benchmark papers and lead in the metric star density by 0.055.

For each safety topic, results vary slightly.
In prompt injection, benchmark papers show higher means of citation density and scientific field count than non-benchmark papers (0.075 vs. 0.046, 3.406 vs. 2.303) but have similar citation count means (20.000 vs. 19.727).
For jailbreak papers, benchmark papers outperform non-benchmark papers in terms of citation density, GitHub star count, and GitHub star density.
However, in hallucination papers, benchmark papers lag behind non-benchmark papers in the means of all metrics except scientific field count, with mean values below 60\% of non-benchmark papers for most metrics.

\begin{figure}[!t]
\centering
\begin{subfigure}{0.45\textwidth}
\centering
\includegraphics[width=0.45\columnwidth]{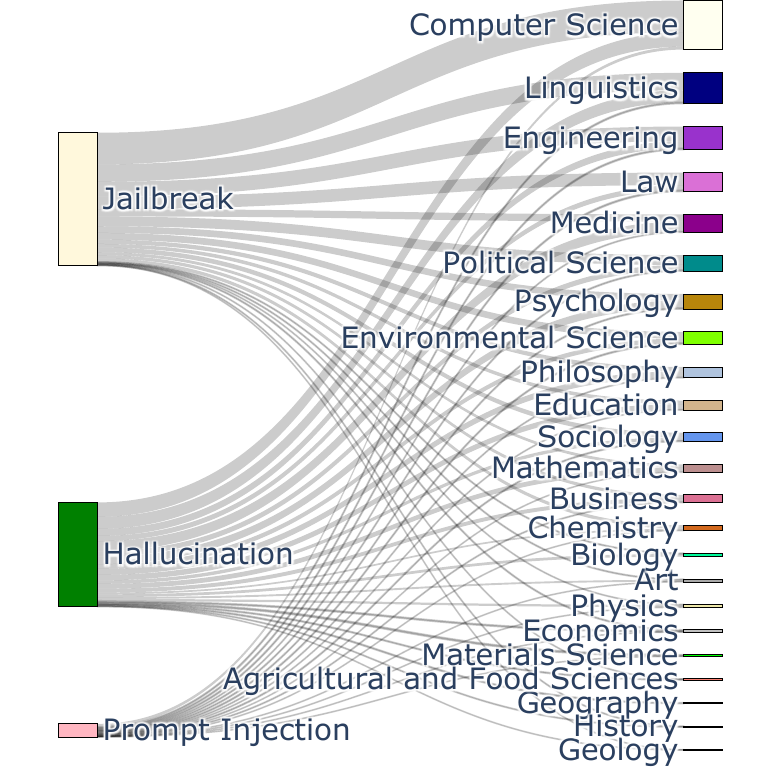}
\caption{
Non-benchmark papers.
}
\label{figure:sankey_other_field}
\end{subfigure}
\begin{subfigure}{0.45\textwidth}
\centering
\includegraphics[width=0.45\columnwidth]{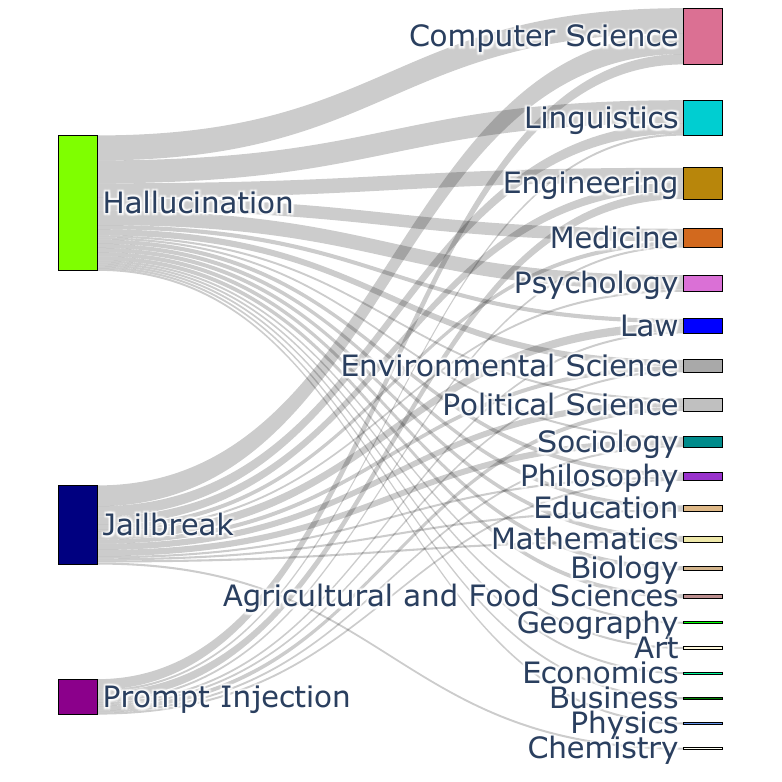}
\caption{
Benchmark papers.
}
\label{figure:sankey_benchmark_field}
\end{subfigure}
\caption{
Distribution of the scientific fields that the LLM safety papers influence.
}
\label{figure:sankey_field}
\end{figure}

In our previous measurements, we find that both non-benchmark and benchmark papers impact an average of over four scientific fields, which indicates that LLM safety papers also hold influence in other domains.
We conduct an in-depth analysis of this aspect, and the distribution of scientific fields is illustrated in~\autoref{figure:sankey_field}.
Considering Computer Science, Linguistics, and Engineering as fields directly related to LLMs, and all others as indirectly related, we find that benchmark papers impact 17 indirectly related fields, while non-benchmark papers affect an additional three indirectly related fields beyond this.
For non-benchmark papers, the most impacted indirectly related field is Law, with 110 such associations.
For benchmark papers, the leading indirectly related field is Medicine, with nine associations.
We also observe that LLM safety papers impact some fields beyond our expected range, such as Arts and Materials Science.
The above results indicate that LLM safety papers have considerable cross-field influence, highlighting the importance of LLM safety research.

\begin{figure}[!ht]
\centering
\includegraphics[width=0.54\columnwidth]{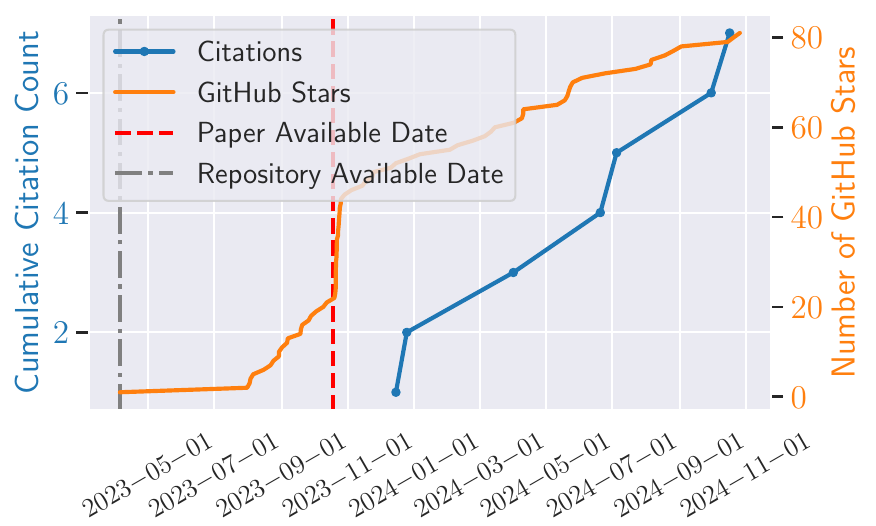}
\caption{
A typical example of the general pattern we identify.
After the paper is publicly available, its GitHub star count increases rapidly.
}
\label{figure:trend}
\end{figure}

We explore the relationship between the cumulative distribution of benchmark papers' citation counts and GitHub star counts over time.
Identifying general patterns is challenging due to variations in the timing of code and paper releases: codes may be released before, after, or simultaneously with the paper.
The consistent pattern observed is that when code is released before the paper, the repository's GitHub star count surges upon the paper's publication.
Of 12 benchmark papers with this release timing, 11 follow this trend (\autoref{figure:trend} of~\autoref{section:supplementary_figures}).
The exception likely reflects the paper's low citation count (only one), limiting community interest in its paper and code.

\section{Descriptive Statistical Analysis of Tool-Based Evaluation}
\label{section:descriptive_code}

\begin{figure*}[!t]
\centering
\begin{subfigure}{0.48\textwidth}
\centering
\includegraphics[width=0.8\textwidth]{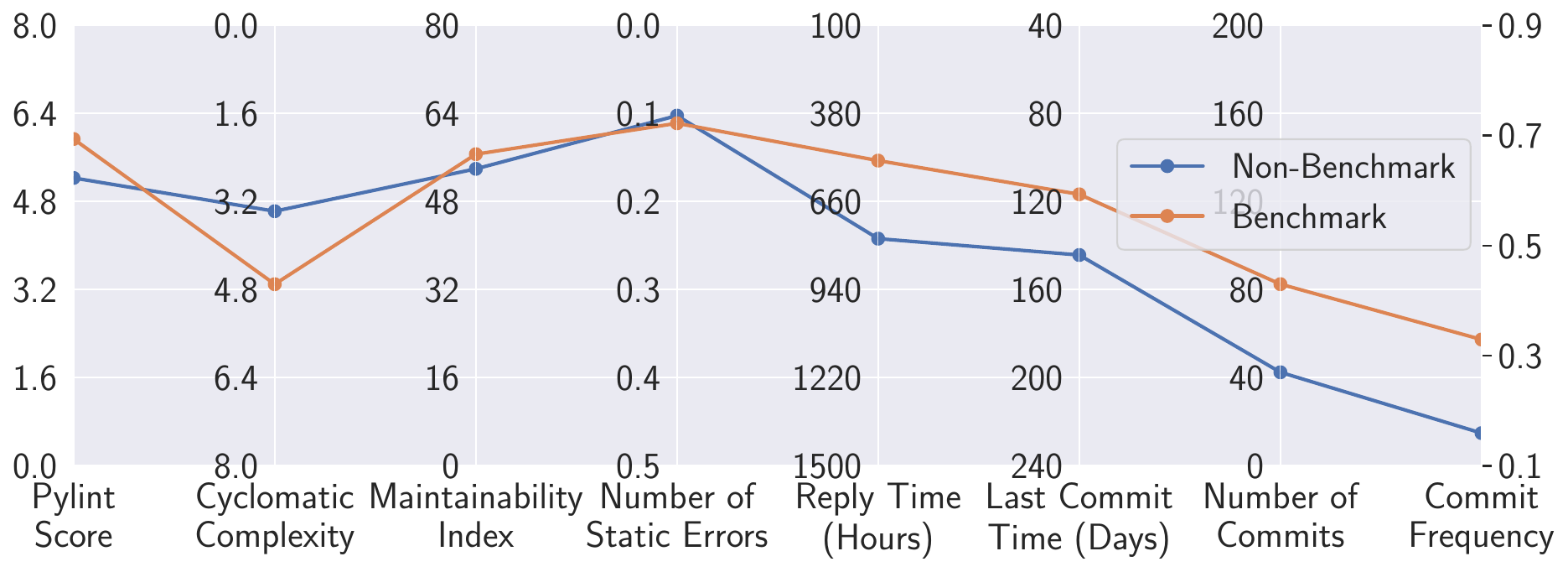}
\caption{
All papers.
}
\label{figure:pax_all_appendix}
\end{subfigure}
\begin{subfigure}{0.48\textwidth}
\centering
\includegraphics[width=0.8\textwidth]{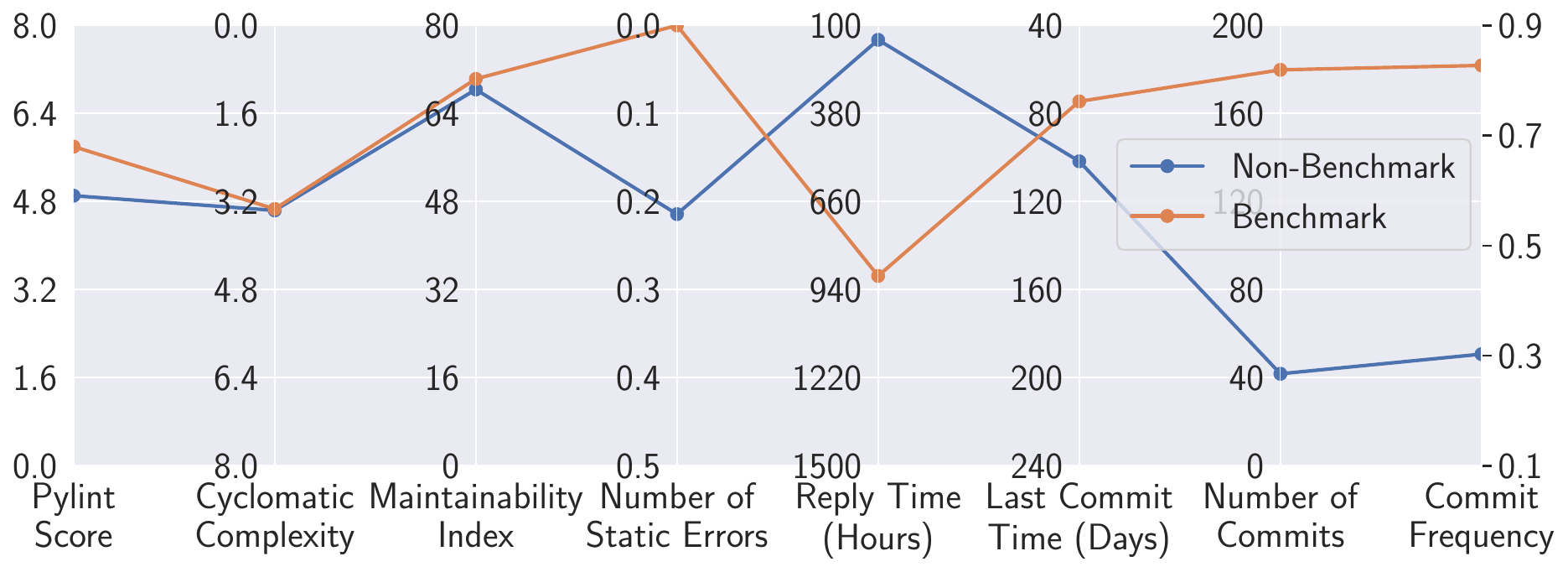}
\caption{
Prompt injection papers.
}
\label{figure:pax_pj}
\end{subfigure}

\begin{subfigure}{0.48\textwidth}
\centering
\includegraphics[width=0.8\textwidth]{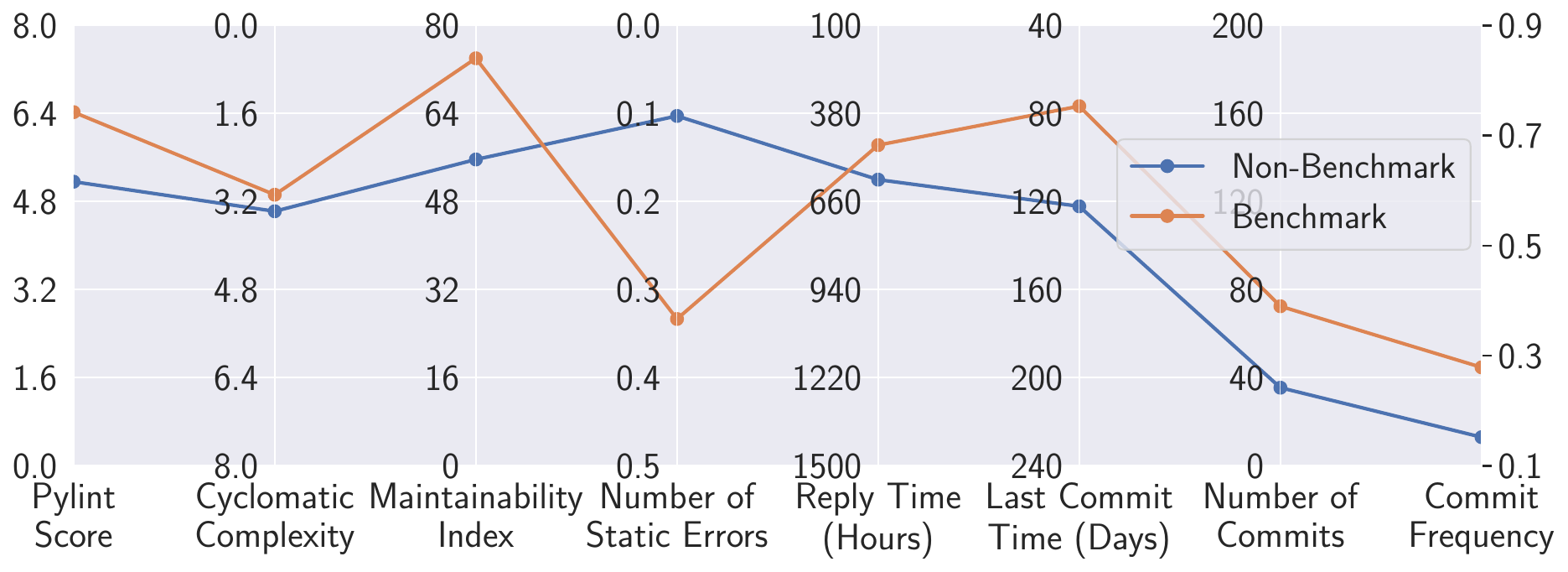}
\caption{
Jailbreak papers.
}
\label{figure:pax_jb}
\end{subfigure}
\begin{subfigure}{0.48\textwidth}
\centering
\includegraphics[width=0.8\textwidth]{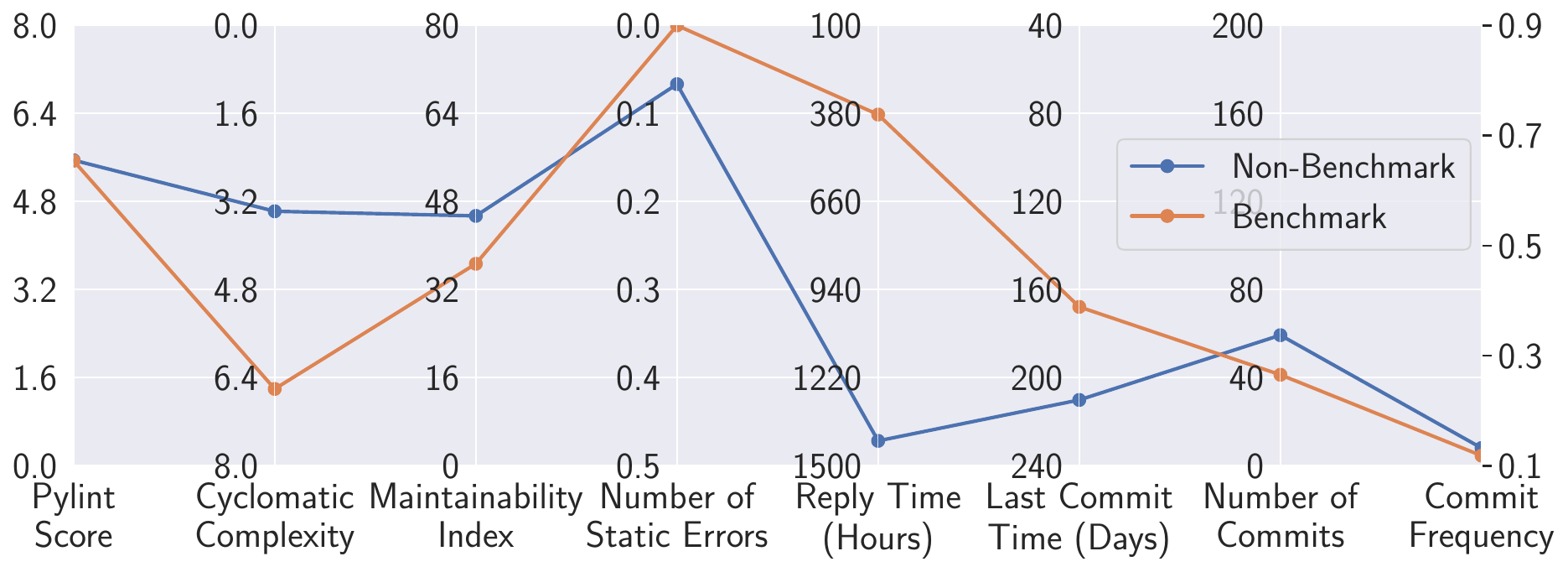}
\caption{
Hallucination papers.
}
\label{figure:pax_halu}
\end{subfigure}
\caption{
Average values of eight metrics related to the code repository quality on benchmark and non-benchmark papers.
We have flipped the axes as needed, so now all points/lines on the upper side indicate better quality or performance.
}
\label{figure:pax_full}
\end{figure*}

\mypara{Code Repository Quality}
The descriptive results of tool-based code quality evaluation are presented in~\autoref{figure:pax_full}.
When considering benchmark papers across all safety topics, we observe that their associated repositories lead or are close to non-benchmark papers' in the mean values of most metrics, with average cyclomatic complexity being the only exception.
Specifically, benchmark papers outperform non-benchmark papers in both average Pylint scores and average maintainability index (5.937 vs. 5.229, 56.606 vs. 53.940).
On the average number of static errors, benchmark papers are quite close to non-benchmark papers (0.111 vs. 0.102).
From the perspective of mean values, benchmark papers outperform non-benchmark papers in all metrics related to code maintenance frequency.
For instance, the average reply time for benchmark papers is significantly shorter than for non-benchmark papers, reduced by 254.204 hours; additionally, the average maintenance frequency of benchmark papers is nearly double that of non-benchmark papers (0.329 vs. 0.159).

A granular analysis by safety topic reveals slightly different observations.
For prompt injection, benchmark papers generally outperform or match non-benchmark papers in the means of code quality metrics, with only a slight lag in cyclomatic complexity means (0.026).
However, their average reply time is significantly longer than non-benchmark ones by 750.683 hours.
In jailbreak, benchmark papers excel in nearly all metric means, notably in the maintainability index (74.032 vs. 55.647).
The exception is the average number of static errors, where benchmark papers have higher (0.333 vs. 0.103), though the absolute value remains low ($<1$), indicating room for improvement.
For hallucination, benchmark papers show similar Pylint average scores but fare worse in average cyclomatic complexity ($\uparrow$3.377) and average maintainability index ($\downarrow$8.699).
Notably, the average reply time for benchmarks is over 1,000 hours shorter than for non-benchmarks.

\section{Exploratory Multiple Linear Regression Analysis}
\label{appendix:regression}

\subsection{Selection of Correlation Analysis Method}
\label{section:analysis_select}

The total number of samples in the Benchmark dataset is relatively limited.
In our study, we included all relevant benchmark works that we retrieved, with a total size of 31.

According to statistical guidelines, approximately ten observations per predictor are needed to obtain stable estimates in regression~\cite{statsols_sample_size_regression,pripp2024_sample_size_prediction_model,newsom_psu_sample_size_power_regression}.
With eight independent variables, a size of 31 provides insufficient degrees of freedom for a robust multivariate analysis.
A Multiple Linear Regression model under these conditions would likely suffer from multicollinearity and unstable parameter estimates.

Instead, the correlation coefficient requires a relatively small sample size (according to statistical guidance, a sample size of 25 or higher is sufficient)~\cite{data_consideration_for_correlation,MD38}.
Therefore, we prioritized correlation coefficients to explore the strength and direction of bivariate associations, which offers a more reliable interpretation given the limited data availability.

Furthermore, we selected Spearman's rank correlation coefficient $\rho$ rather than Pearson's $r$ to mitigate issues related to the distribution of the data.
Compared with Pearson, Spearman has lower data requirements, and its $\rho$ offers greater stability against outliers, which can disproportionately influence results in limited sample sizes~\cite{S04,PS10}.

We assessed statistical significance using permutation tests rather than standard asymptotic methods, which are often ill-suited for small datasets.
By performing 10,000 random shuffles, the permutation test yields exact empirical $p$-values rather than approximations.
This ensures robust inference despite the limited number of observations and the potential non-normality of the underlying population~\cite{G05,EO07,E04}.

Finally, to account for the risk of Type I errors (false positives) arising from repeated testing, we applied the Bonferroni-Holm correction~\cite{H79}.
This step strictly controls the Family-Wise Error Rate (FWER).
To maintain rigorous standards, we only report $\rho$ results with adjusted $p$-values $< 0.05$ as statistically significant in the main text.
Acknowledging that strict FWER control increases the risk of Type II errors (false negatives), we also report $\rho$ results of the raw (unadjusted) $p$-values in the Appendix.
These unadjusted results should be interpreted as exploratory signals requiring further validation in larger cohorts.

\subsection{Exploratory Multiple Linear Regression}

Below, we additionally report two exploratory multiple linear regression models based on ordinary least squares (OLS) fitted on the benchmark dataset (N = 31).
Before performing linear regression, we standardized all predictor variables.
Given the sample size (N = 31) and the relatively large number of predictors (eight per model), the results should be interpreted with caution.
The following regressions are statistically underpowered and numerically unstable.
We therefore include them only as supplementary evidence.

\subsubsection{Regression on Author and Institution Factors}

\noindent\textbf{Dependent variable:} Citation Count \\
\noindent\textbf{Predictors:} Author Number, Institution Number, Area Number, Author H-Index (Top-1), Author Citation Count (Top-1), Institution CSRankings (Top-1), Institution ARWU (Top-1), Search Appearance Frequency

\begin{table}[!ht]
\centering
\caption{OLS Regression Results: Author \& Institution Factors}
\label{table:regression_1}
\scalebox{1.0}{
\setlength{\tabcolsep}{3pt}
\customTableFont
\begin{tabular}{lrrrr}
\toprule
Predictor & Coef. & Std. Err. & t-value & p-value \\
\midrule
Intercept                           & -2.97e-17 & 0.143 & -0.0002 & 1.000 \\
Author Number                       & -0.0802 & 0.172 & -0.466 & 0.646 \\
Institution Number                  & 0.5197  & 0.178 & 2.919 & 0.008 \\
Area Number                         & -0.3347 & 0.182 & -1.839 & 0.079 \\
Author H-Index (Top-1)              & 0.9827 & 0.285 & 3.452 & 0.002 \\
Author Citation Count (Top-1)       & -0.7993 & 0.284 & -2.815 & 0.010 \\
Institution CSRankings (Top-1)      & -0.0246 & 0.175 & -0.141 & 0.889 \\
Institution ARWU (Top-1)            & -0.0306 & 0.180 & -0.170 & 0.867 \\
Search Appearance Frequency         & 0.1658 & 0.169 & 0.978 & 0.339 \\
\midrule
\multicolumn{5}{l}{\textit{Model statistics:}} \\
\multicolumn{5}{l}{R$^{2}$ = 0.549,\quad Adjusted R$^{2}$ = 0.384} \\
\multicolumn{5}{l}{F-statistic = 3.342,\quad p = 0.0118} \\
\multicolumn{5}{l}{Omnibus p = 0.003,\quad JB p = 0.0038,\quad DW = 2.481} \\
\bottomrule
\end{tabular}
}
\end{table}

\noindent\textbf{Notes.}
The model (in~\autoref{table:regression_1}) explains approximately 55\% of the variance (R$^{2}$ = 0.549), but the adjusted R$^{2}$ drops to 0.384 due to the number of predictors relative to the small sample size.
Several coefficients (Author H-Index (Top-1), Institution Number, Author Citation Count (Top-1)) reach nominal significance.
However, the negative coefficient on Author Citation Count (Top-1) is likely caused by multicollinearity and small-N instability.
These results should be treated as exploratory.

\subsubsection{Regression on Code Quality Factors}

\noindent\textbf{Dependent variable:} Citation Count \\
\noindent\textbf{Predictors:} Pylint Score, Cyclomatic Complexity, Maintainability Index, Number of Static Errors, Server Reply Time (Hours), Last Commit Time (Days), Number of Commits, Commit Frequency

\begin{table}[ht!]
\centering
\caption{OLS Regression Results: Code Quality Factors}
\label{table:regression_2}
\scalebox{1.0}{
\setlength{\tabcolsep}{3pt}
\customTableFont
\begin{tabular}{lrrrr}
\toprule
Predictor & Coef. & Std. Err. & t-value & p-value \\
\midrule
Intercept                  & -2.97e-17 & 0.159 & -0.0002 & 1.000 \\
Pylint Score               & -0.1897 & 0.182 & -1.040 & 0.310 \\
Cyclomatic Complexity      & 0.3465 & 0.257 & 1.350 & 0.191 \\
Maintainability Index      & -0.0240 & 0.178 & -0.135 & 0.894 \\
Number of Static Errors    & 0.4453 & 0.165 & 2.699 & 0.013 \\
Server Reply Time (Hours)  & 0.1809 & 0.167 & 1.081 & 0.291 \\
Last Commit Time (Days)    & 0.0045 & 0.291 & 0.015 & 0.988 \\
Number of Commits          & 0.9264 & 0.635 & 1.459 & 0.159 \\
Commit Frequency           & -0.9875 & 0.626 & -1.579 & 0.129 \\
\midrule
\multicolumn{5}{l}{\textit{Model statistics:}} \\
\multicolumn{5}{l}{R$^{2}$ = 0.444,\quad Adjusted R$^{2}$ = 0.242} \\
\multicolumn{5}{l}{F-statistic = 2.195,\quad p = 0.0690} \\
\multicolumn{5}{l}{Omnibus p = 0.019,\quad JB p = 0.0445,\quad DW = 2.240} \\
\bottomrule
\end{tabular}
}
\end{table}

\noindent\textbf{Notes.}
The overall regression (in~\autoref{table:regression_2}) does not reach the 0.05 significance threshold (p = 0.069).
The adjusted R$^{2}$ is only 0.242.
Apart from Number of Static Errors, none of the code-quality metrics exhibits a stable linear association with citation counts under this small-N setting.
These results should therefore be considered exploratory.

\subsection{Summary}

Across both models, the limited sample size (31 observations) combined with eight predictors per model results in high-variance, low-power estimates and unstable coefficients.
These regressions are included for completeness but are not used to support substantive claims.

\section{Execution Time}
\label{section:running_time}

In the main text, the execution time metric for runnable benchmarks is defined as the wall-clock time from the beginning of the debugging process to the first successful completion of the official example script.
Concretely, after cloning the repository, we start timing once we begin to resolve environment dependencies, install required packages, and adjust configuration files or paths as needed.
The timer stops when the example script finishes successfully without errors.
We denote this quantity by
\begin{multline*}
T_{\text{clone} \rightarrow \text{finish}} \;=\; \text{Time from start of debugging} \\
\text{to successful completion of the example script}.
\end{multline*}
This metric is intended to approximate the real user experience: it captures not only the intrinsic runtime of the script, but also the practical effort required to make the benchmark actually work on a fresh machine (e.g., resolving missing dependencies, fixing path issues, or updating outdated instructions).
Providing a concise, lightweight quick-start example that can be executed quickly is generally considered good practice for benchmark usability, and $T_{\text{clone} \rightarrow \text{finish}}$ is designed to reflect this.

In addition, we introduce an alternative metric that explicitly separates environment setup and debugging costs from the intrinsic execution time of the example.
Let $T_{\text{clone} \rightarrow \text{finish}}$ denote the wall-clock time from successful cloning of the repository to the successful completion of the example script, and let $T_{\text{exec}}$ denote the runtime of the example script itself under a correctly configured environment.
We define
\[
T_{\text{setup}} \;=\; T_{\text{clone} \rightarrow \text{finish}} \;-\; T_{\text{exec}}.
\]
By construction, $T_{\text{setup}}$ isolates the overhead introduced by environment setup and debugging while removing confounding factors due to the script's own execution time.

\begin{figure}[!h]
\centering
\begin{subfigure}{0.9\columnwidth}
\centering
\includegraphics[width=0.6\columnwidth]{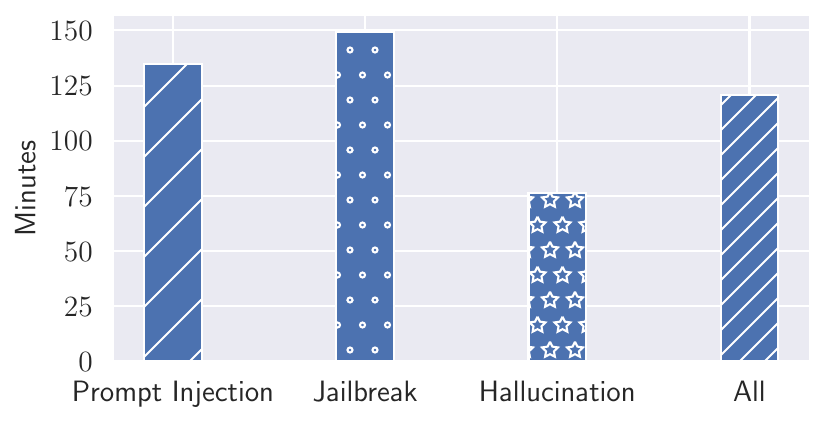}
\caption{
Including script execution time.
}
\label{figure:entire}
\end{subfigure}
\begin{subfigure}{0.9\columnwidth}
\centering
\includegraphics[width=0.6\columnwidth]{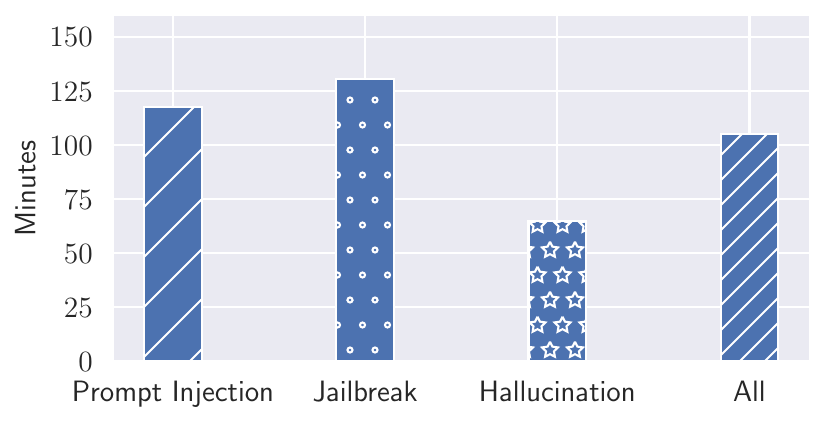}
\caption{
Excluding script execution time.
}
\label{figure:minus}
\end{subfigure}
\caption{
Average time to successfully run the example scripts in the repositories.
}
\label{figure:runnable_demo_time}
\end{figure}

For repositories labeled as \emph{not runnable}, we find that currently all failure arises from environment setup or debugging issues (e.g., missing dependencies, incompatible package versions, broken paths), rather than from excessive demo execution time.
Consequently, whether we adopt the original metric or the alternative definition, the runnability outcome for these repositories remains unchanged.
For the average execution time, we report the results of the two definitions in~\autoref{figure:runnable_demo_time}.

\section{LLM Safety Benchmark Survey}
\label{section:additional_survey}

\subsection*{Section A --- Paper and Repository Influence}

\noindent\textbf{Q1.} In your opinion, which metric is more appropriate for measuring a paper's influence?
\begin{itemize}
    \item Citation count
    \item Citation density
    \item Other: \rule{5cm}{0.1pt}
\end{itemize}

\noindent\textbf{Q2.} In your opinion, which metric is more appropriate for measuring a code repository's influence?
\begin{itemize}
    \item GitHub star count
    \item GitHub star density
    \item Other: \rule{5cm}{0.1pt}
\end{itemize}

\noindent\textbf{Q3.} Do you think an author's prominence affects the influence of a paper or its associated repository?
\begin{itemize}
    \item It affects papers only
    \item It affects repositories only
    \item It affects both
    \item It affects neither
\end{itemize}

\noindent\textbf{Q4.} Do you think the institution behind a work affects the influence of its paper or repository?
\begin{itemize}
    \item It affects papers only
    \item It affects repositories only
    \item It affects both
    \item It affects neither
\end{itemize}

\noindent\textbf{Q5.} Do you think geographical location (e.g., North America / Europe / Asia-Pacific) affects the influence of a paper or repository?
\begin{itemize}
    \item It affects papers only
    \item It affects repositories only
    \item It affects both
    \item It affects neither
\end{itemize}

\noindent\textbf{Q6.} Do you think publication status (e.g., top conference, other conference, preprint) affects a paper's influence?
\begin{itemize}
    \item It affects papers only
    \item It affects repositories only
    \item It affects both
    \item It affects neither
\end{itemize}

\noindent\textbf{Q7.} Do you think public search visibility (e.g., Google Search ranking or frequency) affects the influence of a paper or repository?
\begin{itemize}
    \item It affects papers only
    \item It affects repositories only
    \item It affects both
    \item It affects neither
\end{itemize}

\noindent\textbf{Q8.} What other factors do you believe may influence the visibility or influence of a benchmark work?
\begin{itemize}
    \item Answer: \rule{5cm}{0.1pt}
\end{itemize}

\subsection*{Section B --- Code Quality of Benchmark Repositories}

\noindent\textbf{Q9.} What is the minimum level of quality you expect from a benchmark's code repository?
\begin{itemize}
    \item High quality (almost runnable out-of-the-box)
    \item Basic quality (runnable after some debugging)
    \item No quality requirement as long as it is open-sourced
    \item No need to open-source
    \item Other: \rule{5cm}{0.1pt}
\end{itemize}

\noindent\textbf{Q10.} What type of quality checking do you think benchmark repositories should undergo?
\begin{itemize}
    \item Static analysis
    \item Manual review
    \item Both static analysis and manual review
    \item No quality checks needed
    \item Other: \rule{5cm}{0.1pt}
\end{itemize}

\noindent\textbf{Q11.} When evaluating whether a repository is usable (assuming you are not required to use it), how much time do you typically spend?
\begin{itemize}
    \item Less than 2 hours
    \item 2--4 hours
    \item 4--6 hours
    \item More than 6 hours
\end{itemize}

\noindent\textbf{Q12.} Would you like benchmark repositories to include a minimal runnable example (a minimal test script)?
\begin{itemize}
    \item Yes
    \item Neutral
    \item No
\end{itemize}

\noindent\textbf{Q13.} Which of the following do you think a repository should \emph{at least} include? (multiple choice)
\begin{itemize}
    \item Installation guide
    \item Data guide
    \item Ethical considerations
    \item Other: \rule{5cm}{0.1pt}
\end{itemize}

\noindent\textbf{Q14.} Ideally, which items should the installation or usage guide include? (multiple choice)
\begin{itemize}
    \item Installation guide
    \item Data guide
    \item Ethical considerations
    \item Other: \rule{5cm}{0.1pt}
\end{itemize}

\subsection*{Section C --- Comparison Between Benchmark and Non-Benchmark Repositories}

\noindent\textbf{Q15.} On average, whose code quality do you believe is higher?
\begin{itemize}
    \item Benchmark repositories
    \item Non-benchmark repositories
    \item Not sure: \rule{5cm}{0.1pt}
\end{itemize}

\noindent\textbf{Q16.} On average, whose influence do you believe is higher?
\begin{itemize}
    \item Benchmark papers/repositories
    \item Non-benchmark papers/repositories
    \item Not sure: \rule{5cm}{0.1pt}
\end{itemize}

\subsection*{Section D --- Relationship Between Code Quality and Influence}

\noindent\textbf{Q17.} When you decide whether to cite a paper (assuming no external requirement to reproduce the results), how does code quality affect your decision?
\begin{itemize}
    \item Code quality does not affect whether I cite the paper
    \item Higher code quality increases my willingness to cite
    \item Higher code quality decreases my willingness to cite
    \item Other: \rule{5cm}{0.1pt}
\end{itemize}

\section{Survey Results}
\label{section:survey_results}

We have conducted a 17-question anonymous survey and distributed it via public academic channels ( covering over 50 LLM safety researchers outside our institution).
The survey did not mention this paper or any related context.
We received 17 valid responses during the first round of distribution.
And then we conduct a second round of distribution to collect more responses, and we received 25 additional responses.
In total, we received 42 valid responses.

\subsection{Paper and Repository Influence (Q1--Q8)}

\mypara{Paper influence}
25/42 respondents considered \textit{citation density} to be the most appropriate metric.
3/42 suggested ``other metrics,'' but none provided concrete alternatives.

\mypara{Repository influence}
For code repositories, 30/42 preferred \textit{GitHub star density}.
5/42 noted that for \textit{LLM models} (not applicable in this study), HuggingFace statistics can sometimes be more informative.

\mypara{Author, institution, geolocation}
For Q3--Q5, most respondents agreed that author prominence, institutional affiliation, and geolocation affect the visibility or influence of papers and repositories.

\mypara{Publication status}
Responses to Q6 were divided: 17/42 believed publication venue matters, whereas 21/42 believed it does not.

\mypara{Search visibility}
For Q7, 23/42 stated that public search visibility affects \textit{papers only}, while 8/42 believed it affects \textit{both} papers and repositories.

\mypara{Additional factors}
In Q8, four respondents highlighted that blog posts from major labs such as OpenAI or Anthropic can substantially influence visibility.

\subsection{Code Repository Quality (Q9--Q14)}

\mypara{Minimum acceptable standard (Q9)}
36/42 selected ``basic quality'' (runnable after some debugging).
One respondent commented that they would prefer \emph{no release} over a poorly maintained open-source repository.

\mypara{Quality checking (Q10)}
33/42 preferred \textit{manual review}, and 14/42 preferred \textit{manual review + static analysis}.
No respondent favored static analysis alone.

\mypara{Time acceptable for usability checking (Q11)}
32/42 were willing to spend $<2$ hours; 10/42 accepted 2--4 hours; no respondent accepted more than 6 hours.

\mypara{Minimal runnable example (Q12)}
38/42 wanted a minimal runnable example; 3/42 were neutral.

\mypara{Minimum required repository contents (Q13)}
34/42 selected the installation guide only.
14/42 selected both installation and data guides.
This indicates that installation instructions are considered the baseline component.

\mypara{Ideal repository contents (Q14)}
All 42/42 selected both installation and data guides as ideal requirements.
6/42 additionally recommended including ethical or responsible-use considerations.

\subsection{Benchmark vs. Non-Benchmark Comparison (Q15--Q16)}

\mypara{Code quality (Q15)}
36/42 believed benchmark repositories have higher code quality.

\mypara{Influence (Q16)}
Responses were mixed: 17/42 selected benchmark work, 21/42 selected non-benchmark work, and 4/42 were unsure.
3 of the ``unsure'' respondents noted that non-benchmark work has a ``higher upper bound but also a very low lower bound.''

\subsection{Relationship Between Code Quality and Influence (Q17)}

20/42 reported that higher code quality increases their willingness to cite a paper.
14/42 stated that code quality does not affect their citation decision.
No respondent reported that higher code quality decreases their willingness to cite.

\section{Supplementary Tables and Figures}
\label{section:supplementary_figures}

\begin{table}[!ht]
  \centering
  \caption{Summary of the metrics used in adoption evaluation.
  }
  \label{table:influence_metrics}
  \scalebox{0.77}{
    \setlength{\tabcolsep}{3pt}
    \customTableFont
    \begin{tabular}{m{0.3\columnwidth}|m{0.3\columnwidth}}
    \toprule
    \makecell{\textbf{Dimension}} & \makecell{\textbf{Metric}} \\
    \midrule
    \makecell{\multirow{2}[0]{*}[0ex]{\makecell{Academic Community}}} & \makecell{Citation Count} \\
          & \makecell{Citation Density} \\
    \midrule
    \makecell{\multirow{2}[0]{*}[0ex]{\makecell{Open-Source\\Community}}} & \makecell{GitHub Star Count} \\
          & \makecell{GitHub Star Density}  \\
    \midrule
    \makecell{\multirow{1}[0]{*}[0ex]{\makecell{Cross-Disciplinary}}} & \makecell{Scientific Field Count} \\
    \bottomrule
    \end{tabular}
    }
\end{table}

\begin{figure}[!ht]
\centering
\includegraphics[width=0.54\columnwidth]{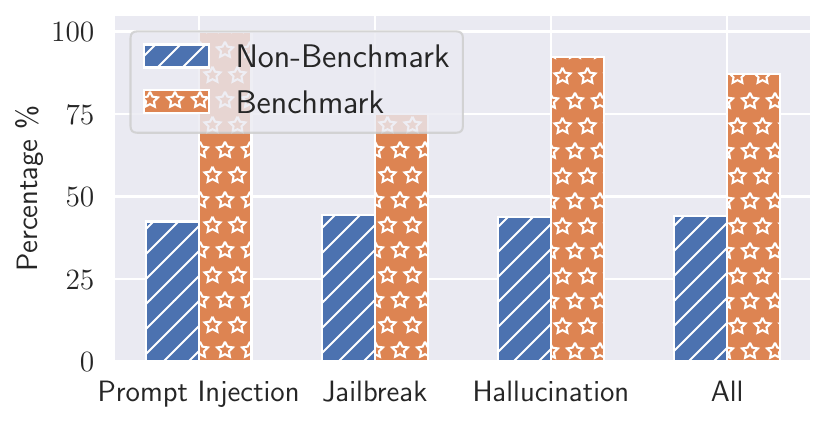}
\caption{
GitHub repository availability proportions.
}
\label{figure:github_availability}
\end{figure}

\begin{figure}[!ht]
    \centering
    \includegraphics[width=0.54\columnwidth]{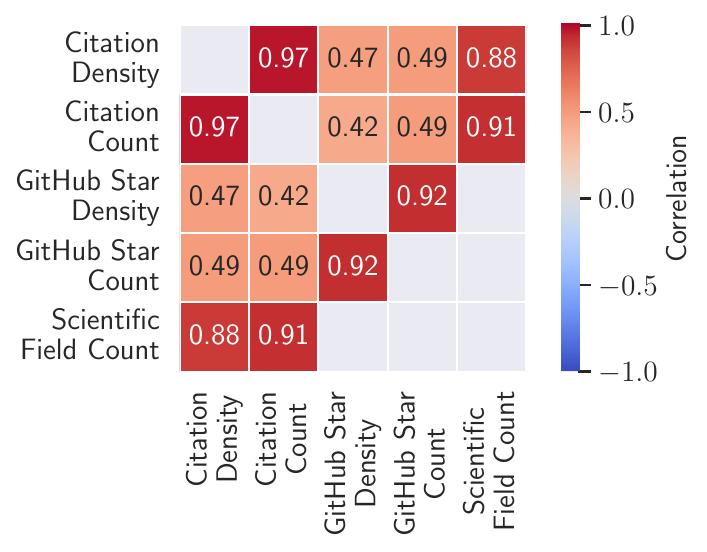}
    \caption{
        Spearman correlation $\rho$ matrix of the adoption metrics (those with \( p \geq 0.05 \) are omitted).
    }
    \label{figure:relationship_influence}
\end{figure}

\begin{figure}[!ht]
\centering
\begin{subfigure}{0.9\columnwidth}
\centering
\includegraphics[width=0.6\columnwidth]{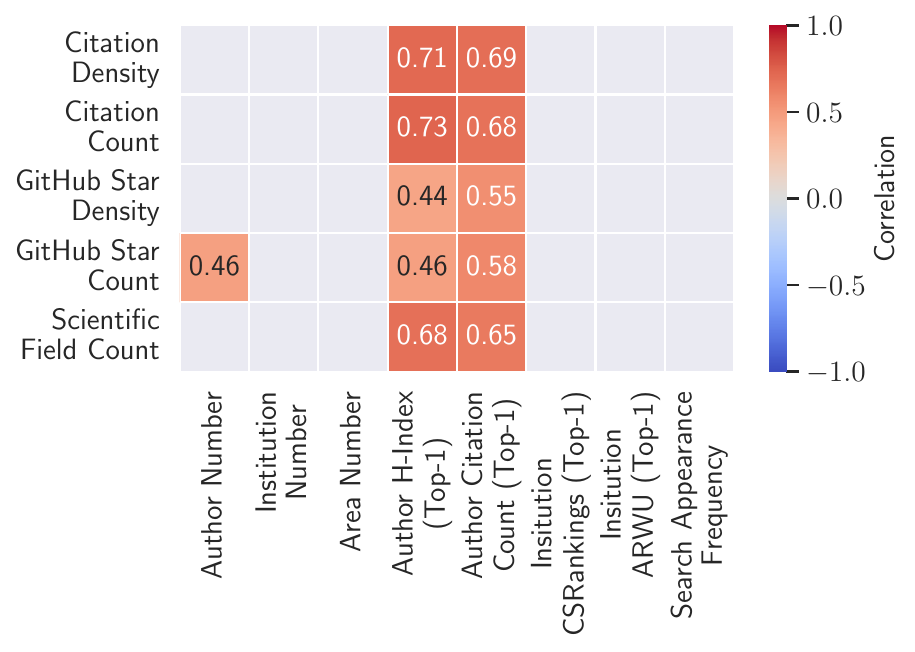}
\caption{
$\rho$ matrix (raw $p<0.05$).
}
\end{subfigure}
\begin{subfigure}{0.9\columnwidth}
\centering
\includegraphics[width=0.6\columnwidth]{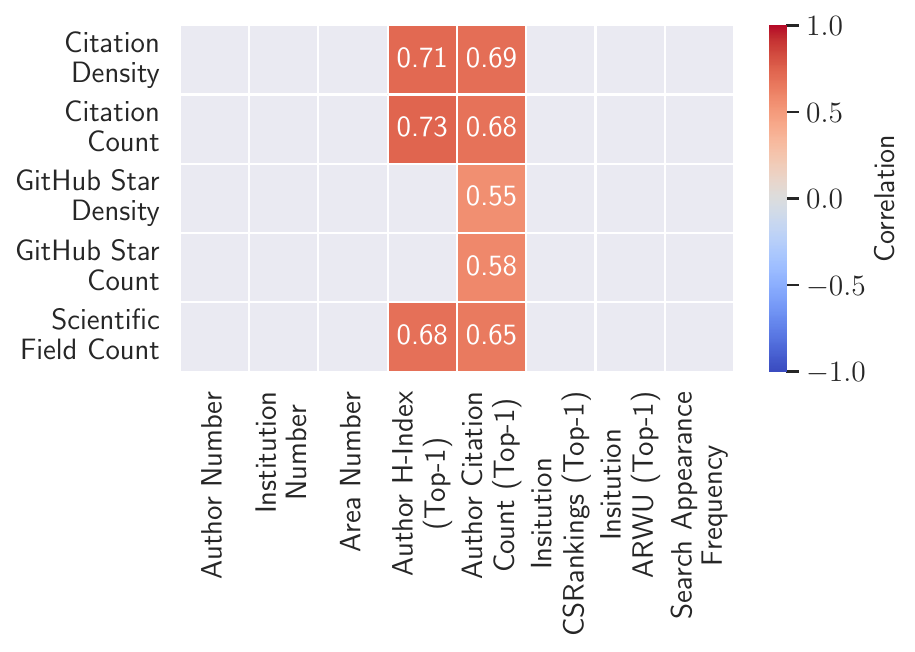}
\caption{
$\rho$ matrix (adjusted $p<0.05$).
}
\end{subfigure}
\caption{
Spearman correlation matrices between the adoption metrics and the potential quantitative factors.
The unadjusted p-values on the left can be interpreted exploratively.
}
\label{figure:relationship_influence_other}
\end{figure}

\begin{figure}[!ht]
\centering
\begin{subfigure}{0.9\columnwidth}
\centering
\includegraphics[width=0.6\columnwidth]{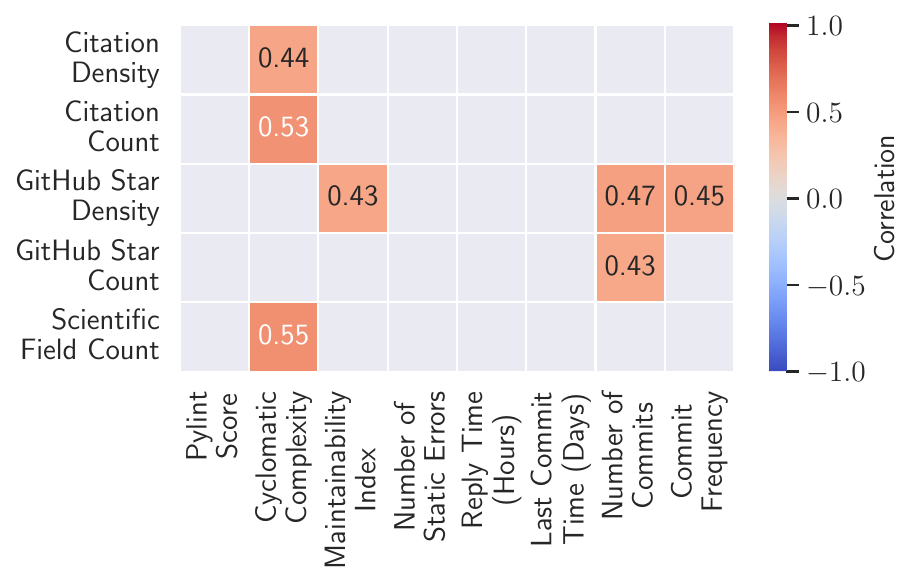}
\caption{
$\rho$ matrix (raw $p<0.05$).
}
\end{subfigure}
\begin{subfigure}{0.9\columnwidth}
\centering
\includegraphics[width=0.6\columnwidth]{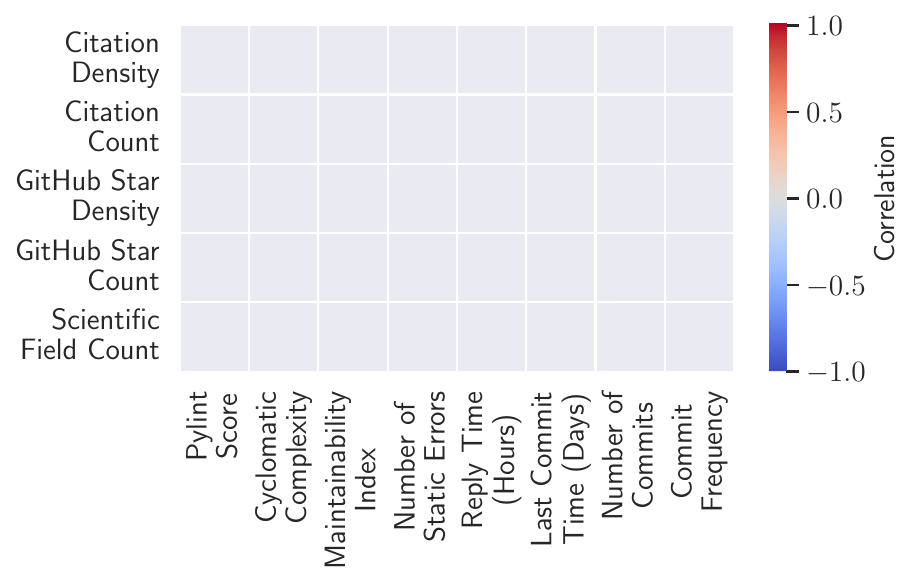}
\caption{
$\rho$ matrix (adjusted $p<0.05$).
}
\end{subfigure}
\caption{
Spearman correlation matrices between the adoption metrics and the code repository quality metrics.
The unadjusted p-values on the left can be interpreted exploratively.
}
\label{figure:relationship_influence_code}
\end{figure}

\begin{figure}[!ht]
\centering
\begin{subfigure}{0.9\columnwidth}
\centering
\includegraphics[width=0.6\columnwidth]{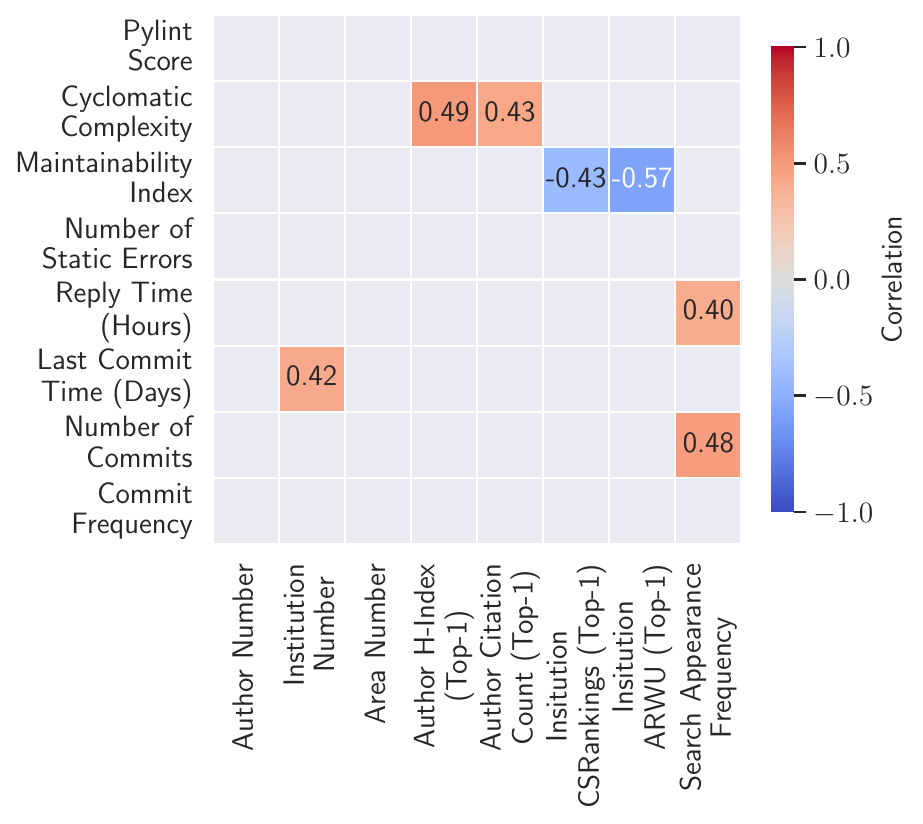}
\caption{
$\rho$ matrix (raw $p<0.05$).
}
\end{subfigure}
\begin{subfigure}{0.9\columnwidth}
\centering
\includegraphics[width=0.6\columnwidth]{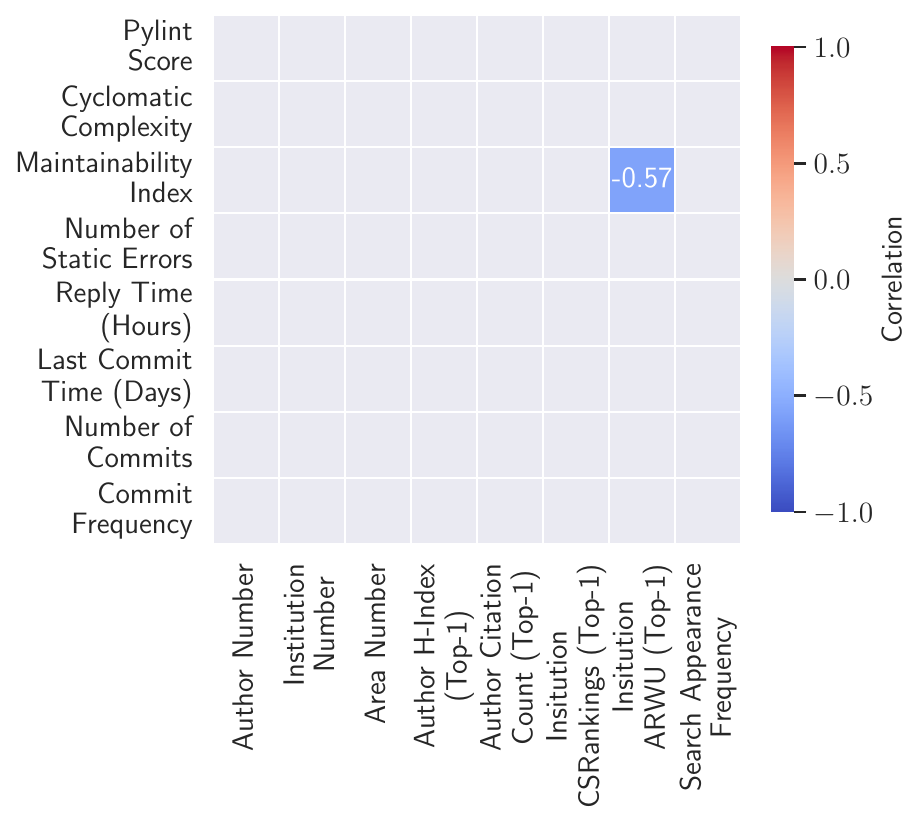}
\caption{
$\rho$ matrix (adjusted $p<0.05$).
}
\end{subfigure}
\caption{
Spearman correlation matrices between the code repository quality metrics and the potential quantitative factors.
The unadjusted p-values on the left can be interpreted exploratively.
}
\label{figure:relationship_code_other}
\end{figure}

\begin{figure}[!ht]
\centering
\begin{subfigure}{0.48\columnwidth}
\centering
\includegraphics[width=0.7\columnwidth]{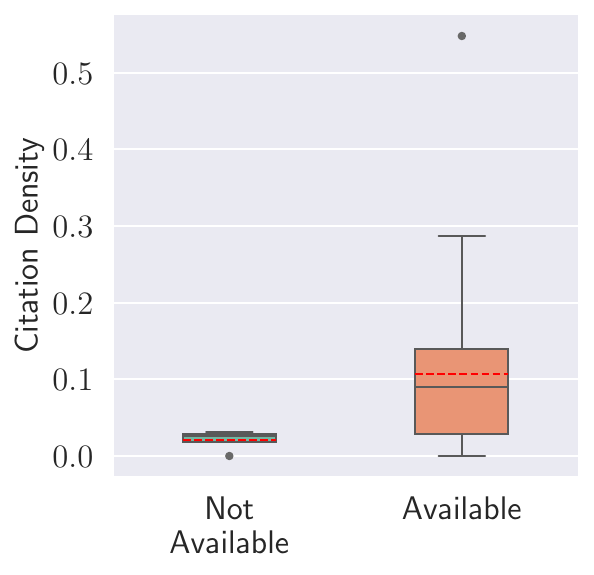}
\caption{
Code availability.
}
\end{subfigure}
\begin{subfigure}{0.48\columnwidth}
\centering
\includegraphics[width=0.7\columnwidth]{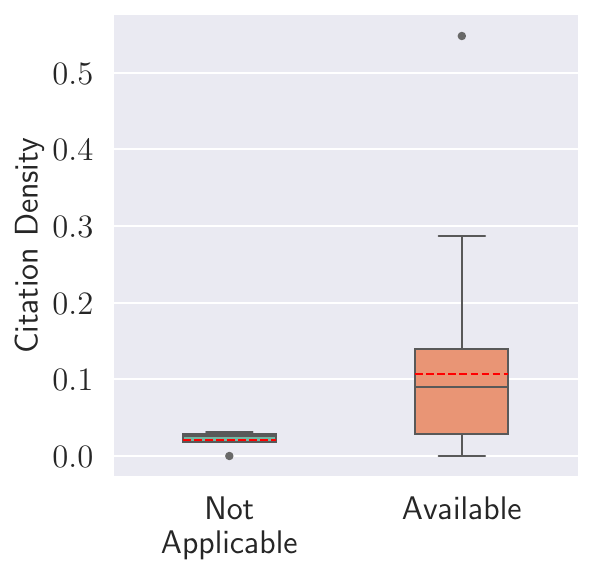}
\caption{
Data availability.
}
\end{subfigure}
\caption{
Box plots of citation density by group.
The red dashed lines represent the means.
}
\label{figure:influence_availability}
\end{figure}

\begin{figure}[!ht]
\centering
\begin{subfigure}{0.48\columnwidth}
\centering
\includegraphics[width=0.7\columnwidth]{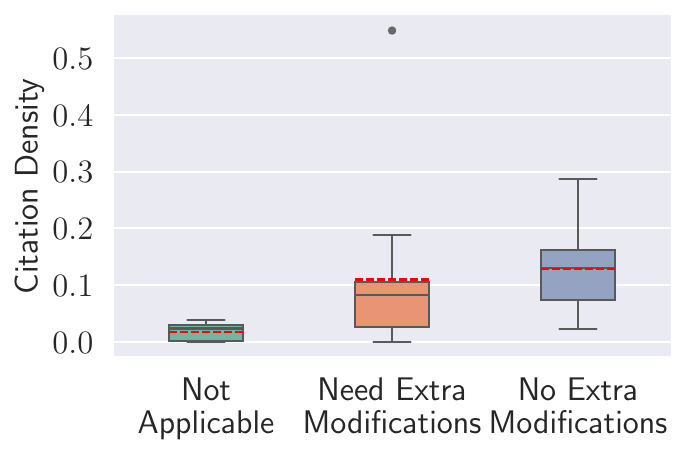}
\caption{
Modification status.
}
\end{subfigure}
\begin{subfigure}{0.48\columnwidth}
\centering
\includegraphics[width=0.7\columnwidth]{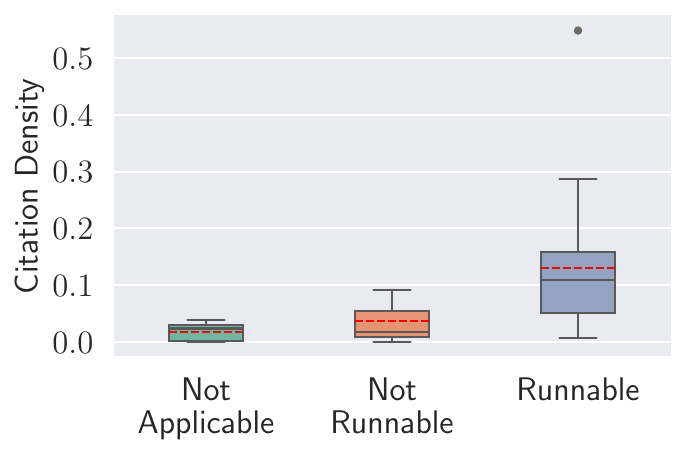}
\caption{
Runnable code status.
}
\end{subfigure}
\caption{
Box plots of citation density and the status of extra modifications and runnable code.
The red dashed lines represent the mean values.
}
\label{figure:influence_runnable_bugs}
\end{figure}

\begin{figure}[!ht]
\centering
\begin{subfigure}{0.32\textwidth}
\centering
\includegraphics[width=0.5\textwidth]{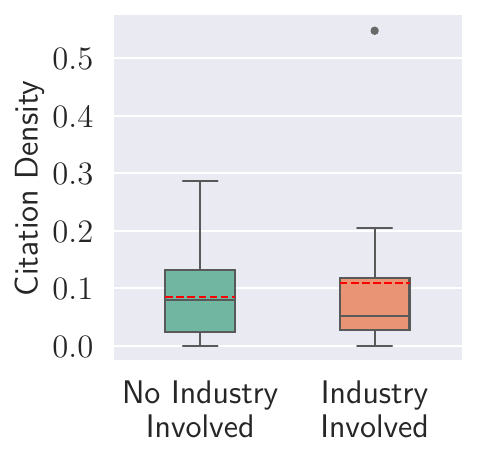}
\caption{
Industry involvement.
}
\label{figure:boxwithmean_industry}
\end{subfigure}

\begin{subfigure}{0.32\textwidth}
\centering
\includegraphics[width=0.5\textwidth]{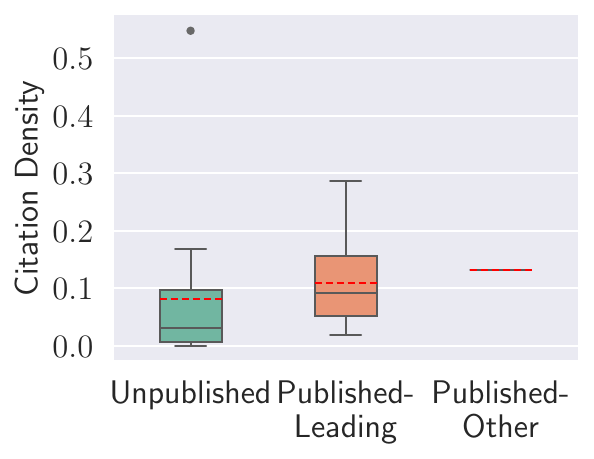}
\caption{
Publication.
}
\label{figure:boxwithmean_topic}
\end{subfigure}

\begin{subfigure}{0.32\textwidth}
\centering
\includegraphics[width=0.5\textwidth]{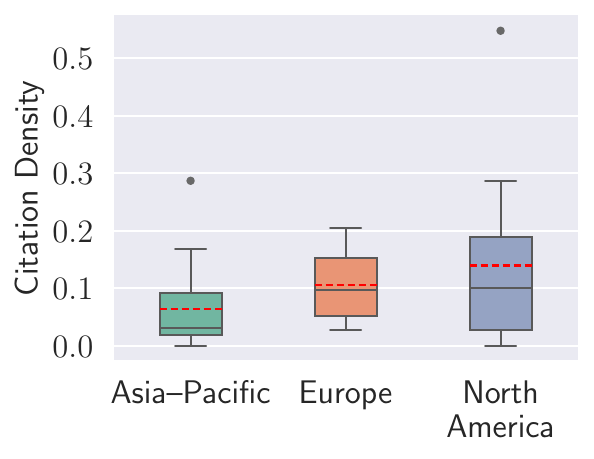}
\caption{
Area.
}
\label{figure:boxwithmean_big_area}
\end{subfigure}
\caption{
Box plots of citation density and various potential qualitative factors.
The red dashed lines in the box plots represent the mean values.
The further to the right the x-axis tick is, the higher the mean value of the corresponding box.
}
\label{figure:boxwithmean}
\end{figure}

\begin{figure}[!ht]
\centering
\begin{subfigure}{0.32\textwidth}
\centering
\includegraphics[width=0.5\textwidth]{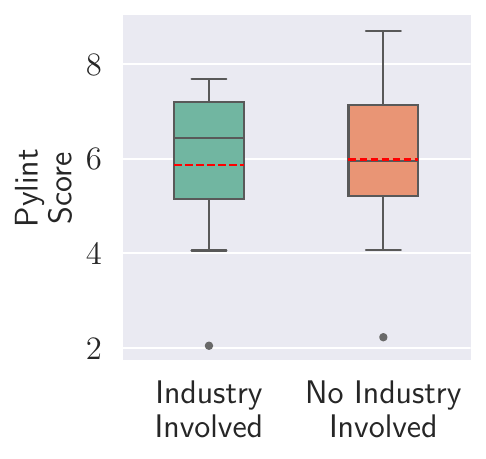}
\caption{
Industry involvement.
}
\end{subfigure}
\begin{subfigure}{0.32\textwidth}
\centering
\includegraphics[width=0.5\textwidth]{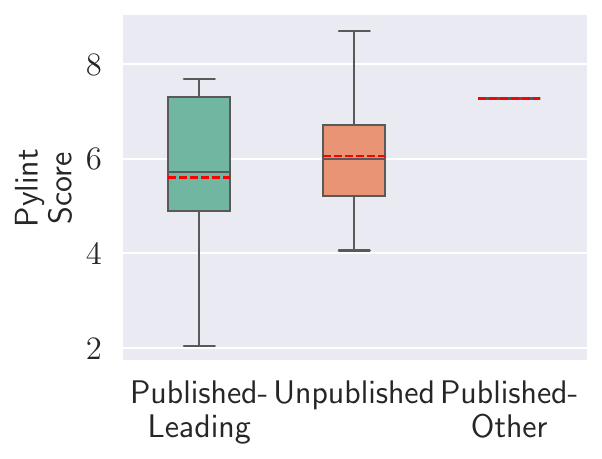}
\caption{
Publication.
}
\end{subfigure}
\begin{subfigure}{0.32\textwidth}
\centering
\includegraphics[width=0.5\textwidth]{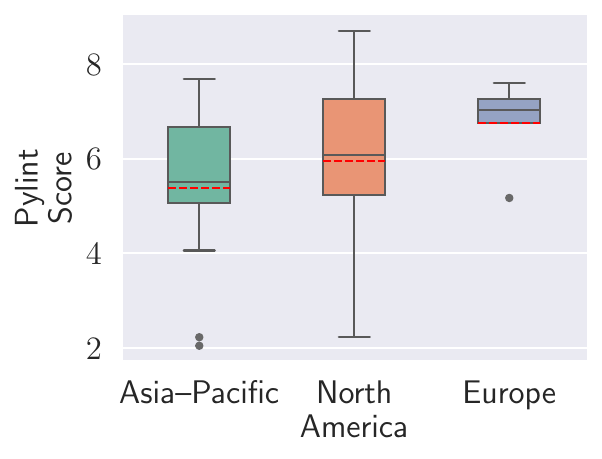}
\caption{
Area.
}
\end{subfigure}
\caption{
Box plots of Pylint score and various potential qualitative factors.
The red dashed lines in the box plots represent the mean values.
The further to the right the x-axis tick is, the higher the mean value of the corresponding box.
}
\label{figure:boxwithmean_pylint}
\end{figure}

\end{document}